\documentclass[final,3p,times,onecolumn]{elsarticle}

\usepackage{subfig}
\usepackage{amsmath, bm}
\usepackage{upgreek}
\usepackage{multirow}
\usepackage{siunitx}
\usepackage[export]{adjustbox}
\usepackage{booktabs}
\usepackage{scalerel}

\usepackage[hidelinks]{hyperref}
\usepackage[nameinlink, capitalise]{cleveref}

\Crefname{figure}{Fig.}{Figs.}
\Crefname{equation}{Eq.}{Eqs.}

\makeatletter
\def\ps@pprintTitle{%
  \let\@oddhead\@empty
  \let\@evenhead\@empty
  \let\@oddfoot\@empty
  \let\@evenfoot\@oddfoot
}
\makeatother

\begin{document}

\begin{frontmatter}

\title{Physically recurrent neural network for rate and path-dependent heterogeneous materials in a finite strain framework}

\author[a]{M. A. Maia} 
\author[a]{I. B. C. M. Rocha}
\author[a]{D. Kova\v cevi\' c}
\author[a]{F. P. van der Meer}

\address[a]{{Delft University of Technology, Department of Civil Engineering and Geosciences}, {PO Box 5048}, 2600 GA, Delft, {The Netherlands}}

\begin{abstract}
In this work, a hybrid physics-based data-driven surrogate model for the microscale analysis of heterogeneous material is investigated. The proposed model benefits from the physics-based knowledge contained in the constitutive models used in the full-order micromodel by embedding them in a neural network. Following previous developments, this paper extends the applicability of the physically recurrent neural network (PRNN) by introducing an architecture suitable for rate-dependent materials in a finite strain framework. In this model, the homogenized deformation gradient of the micromodel is encoded into a set of deformation gradients serving as input to the embedded constitutive models. These constitutive models compute stresses, which are combined in a decoder to predict the homogenized stress, such that the internal variables of the history-dependent constitutive models naturally provide physics-based memory for the network. To demonstrate the capabilities of the surrogate model, we consider a unidirectional composite micromodel with transversely isotropic elastic fibers and elasto-viscoplastic matrix material. The extrapolation properties of the surrogate model trained to replace such micromodel are tested on loading scenarios unseen during training, ranging from different strain-rates to cyclic loading and relaxation. Speed-ups of three orders of magnitude with respect to the runtime of the original micromodel are obtained.

\end{abstract}

\begin{keyword}
Neural networks \sep Heterogeneous materials \sep Rate-dependency \sep Path-dependency \sep Constitutive model

\end{keyword}

\end{frontmatter}

\newcommand{\fesqr}{FE$^2$}
\newcommand{\pd}[2]{\frac{\partial{#1}}{\partial{#2}}}
\newcommand{\strainvec}{\bm{\varepsilon}}
\newcommand{\stressrot}{\bm{\sigma}_{\textrm{F}}}
\newcommand{\predstressrot}{\widehat{\bm{\sigma}}^{\Omega}_{\textrm{F}}}
\newcommand{\strainstresscurve}{\bm{\varepsilon}\text{-}\bm{\sigma}}
\newcommand{\intvarvec}{\bm{\alpha}}
\newcommand{\macrostrainvec}{\bm{\varepsilon}^{\Omega}}
\newcommand{\macrostressvec}{\bm{\sigma}^{\Omega}}
\newcommand{\microstressvec}{\bm{\sigma}^{\omega}}
\newcommand{\microstrainvec}{\bm{\varepsilon}^{\omega}}
\newcommand{\macrodispfield}{\mathbf{u}^\Omega}
\newcommand{\microdispfield}{\mathbf{u}^\omega}
\newcommand{\dispfield}{\mathbf{u}}
\newcommand{\weights}{\mathbf{W}}
\newcommand{\Wjmat}{\mathbf{W}_{j}}
\newcommand{\Ljmat}{\mathbf{W}_j}
\newcommand{\biases}{\mathbf{b}}
\newcommand{\hidden}{\mathbf{h}}
\newcommand{\histunits}{\mathbf{h}}
\newcommand{\stiffmatrix}{\mathbf{D}}
\newcommand{\xvec}{\mathbf{x}}
\newcommand{\Bmat}{\mathbf{B}}
\newcommand{\macrodmat}{\mathbf{D}^{\Omega}}
\newcommand{\Ftensor}{\mathbf{F}}
\newcommand{\Rtensor}{\mathbf{R}}
\newcommand{\Utensor}{\mathbf{U}}
\newcommand{\predstressunrot}{\widehat{\boldsymbol{\sigma}}_\text{U}^{\Omega}}
\newcommand{\stressunrot}{\boldsymbol{\sigma}_\text{U}}
\newcommand{\Umacrotensor}{\mathbf{U}^{\Omega}}
\newcommand{\Umicrotensor}{\mathbf{U}^{\omega}}
\newcommand{\Bmacrotensor}{\mathbf{B}^{\Omega}}
\newcommand{\Bmicrotensor}{\mathbf{B}^{\omega}}
\newcommand{\Idtensor}{\mathbf{I}}
\newcommand{\Fmacrotensor}{\mathbf{F}^{\Omega}}
\newcommand{\Fmicrotensor}{\mathbf{F}^{\omega}}
\newcommand{\Rmacrotensor}{\mathbf{R}^{\Omega}}
\newcommand{\incdefgrad}{\Delta\mathbf{F}}
\newcommand{\defgrad}[2][]{\mathbf{F}_{#1}^{#2}}
\newcommand{\stress}[2][]{\boldsymbol{\sigma}_{#1}^{#2}}
\newcommand{\intvar}[2][]{\boldsymbol{\alpha}_{#1}^{#2}}
\newcommand{\predmacrostress}{\widehat{\boldsymbol{\sigma}}^{\Omega}_t}
\newcommand{\norm}[1]{\left\lVert#1\right\rVert}

\newcommand{\testmono}{\mathcal{T}_\text{mono}}
\newcommand{\testslower}{\mathcal{T}_{\text{slower}}}
\newcommand{\testfaster}{\mathcal{T}_{\text{faster}}}
\newcommand{\testunlfixed}{\mathcal{T}_{\text{unl}}^{\text{fixed}}}
\newcommand{\testunlgpbased}{\mathcal{T}_{\text{unl}}^{\text{prop.\,GP}}}
\newcommand{\testunlgp}{\mathcal{T}_{\text{unl}}^{\text{non--prop.\,GP}}}

\section{Introduction}
\label{sec:intro}

Owing to their high flexibility and potential to reduce computational costs, machine learning techniques are increasingly popular in solid mechanics. These techniques can be especially useful in micromechanical and multiscale analysis, where the accurate representations of complex materials are often compromised by the notoriously high computational costs associated with these methods. Complex heterogeneous materials can be modeled on a lower scale, the microscale, through a so-called Representative Element Volume (RVE), a micromodel assumed to statistically represent the material behavior. At that scale, classical constitutive models can be conveniently employed to describe the behavior of each of the constituents. This allows for a accurate representation of intricate phenomena in the composite material behavior without the need for assumptions about the macroscopic material behavior. The generality of the method, however, is not without trade-offs. Large micromodels, fine meshes, and path and rate-dependent materials are some of the features that result in exceedingly high computational costs.

A common approach to tackle this issue is to replace the micromodel altogether with a surrogate model that reproduces the relation between the homogenized strains and stresses at a lower computational cost. In applications involving path-dependent materials, variations of Recurrent Neural Networks (RNN) are the most popular choice , but other surrogate modeling strategies built on Gaussian Processes (GPs) and dimensionality reduction techniques (\textit{e.g.} Proper Orthogonal Decomposition (POD) and Hyper-reduction methods) have also showed potential for reducing the computational costs \cite{OLIVER2017,GHAVAMIAN2017,ROCHA2019,ROCHA2021}. 

When it comes to neural networks, the more complex architectures derived from RNNs, such as Gated Recurrent Unit (GRU) and Long Short-Term Memory (LSTM), are the predominant choice at present. These models can handle sequential data through mechanisms that propagate information from previous to later states when processing a sequence (\textit{e.g.} an entire path of $\varepsilon$-$\sigma$ pairs). Several works showcase their potential in modeling path-dependent behavior 
for both homogeneous \cite{HEIDER2020} and heterogeneous materials \cite{WU2022114476, LOGARZO2021, GORJI2020, Mozaffar26414}. Nevertheless, several unresolved issues and challenges remain. One of them lies in the fact that in spite of the similarity between the role of the hidden state in RNNs and the internal variables in a constitutive model, the network mechanism is still regarded as a black-box and insights into any latent physical patterns learned by the network are thus far limited to simple settings (\textit{e.g.} homogeneous material in 1D problems \cite{Koeppe2021, LIU2023105329}). 

Another pressing issue in these networks is their limited ability to extrapolate. This is usually tackled with ever larger training sets and intricate design of experiments that aim to uniformly/densely cover the space of strain paths. A complicating aspect is the curse of dimensionality. In that regard, frameworks based on RNNs and variations are typically exemplified in 1D or 2D problems, but even in those cases a large variety of loading/unloading cases is required to cover similar paths and patterns encountered in actual microscale simulations. Recent works \cite{ghane2023recurrent,CHEUNG2024} illustrate the hurdles with predicting loading types different from the ones used for training. In \cite{ghane2023recurrent}, a strategy based on transfer learning is employed to improve the training performance of LSTMs and GRUs and overcome feature sparsity issues. For this, the authors train a network on data generated using a random walk strategy, and then use the optimized parameters in the initialization of a second network trained to predict cyclic loading. \citet{CHEUNG2024} extend that idea to train GRUs with multi-fidelity data, helping reduce the computational cost of generating large high-fidelity training datasets. 

An alternative approach to uncover the black-box nature of these methods is to introduce physics knowledge into the ML-based model. Following that philosophy, Physics-Informed Neural Networks (PINNs) are likely the biggest exponent. Although these networks have been initially designed to solve partial differential equations, the idea of enriching the loss function with extra terms to enforce physics constraints has quickly found its way into the material modeling community. For instance, to predict displacement and stress fields, in addition to terms corresponding to the Neumann and Dirichlet boundary conditions in the loss function of a PINN, one could also include physics-based constraints such as Karush-Kuhn-Tucker conditions when modeling plasticity, as done by \citet{HAGHIGHAT2021} or the evolution of the plastic strain-rate for viscoplastic materials, as shown in \cite{Arora2022}. In spite of the additional information, \citet{HAGHIGHAT2021} report no benefit in using PINNs as forward solvers and \citet{Arora2022} comment on the degrading performance in extrapolation. 

Another way to leverage physical consistency in NNs is to encode the physical knowledge directly in the architecture design \cite{Masi2021, EGHBALIAN2023, garanger2023symmetryenforcing}. For instance, in \cite{Masi2021}, the authors proposed a framework where stresses and dissipation are obtained through the differentation of the learned energy potential function. Strategic architectural choices can also help enforce a specific behavior, as done in \cite{garanger2023symmetryenforcing}. In that work, tensor-based features and activation functions are used in feed-forward and GRUs to enforce material symmetries.   

When dealing with materials with time-dependency, however, the extra dimensionality related to strain-rate sensitivity adds a new depth to the problem. In some works, the strain-rate \cite{WEN2021, Bhattacharya2023} and/or the time increment \cite{Ge2021} have been explicitly included in the feature space. In others, a fixed time increment is considered \cite{Ghavamian2019, Chen2021}. More recently, \citet{Eghtesad2023} proposed a framework based on the dual potential function to describe rate-dependent viscoplastic flow response in metals. The authors take advantage of input-convex NNs to enforce thermodynamic consistency and leverage automatic differentation to compute gradients of the output with respect to the inputs, which are used for solving the implicit time-stepping algorithm employed in their elasto-viscoplastic model. Nevertheless, the method is not suitable for FE simulations yet as arbitrary loading and boundary conditions can take place and only uniaxial deformations were considered. 

In all of these works, to train a surrogate for rate-dependence the training data needs to account not only for a good coverage of strains but also strain-rates. To deal with time-dependency in a more seamless manner, we propose to expand the applicability of the approach presented in \cite{Maiaetal2023}, namely the Physically Recurrent Neural Network (PRNN). In that work, a network with embedded physics-based material models was used to accelerate multiscale analysis of path-dependent heterogeneous materials. The core idea is that the macroscopic strain can be encoded into a set of strains for (fictitious) material points, from which the stress is computed using the same material models and properties as in the micromodel. With these stresses, a decoder is applied to obtain the macroscopic stresses in a homogenization-like step. 

A key element in the proposed architecture consists of letting the material model that evaluates the fictitious material point stress also handle the evolution of its own internal variables. This way, the network inherits the assumptions built into the material models used in the micromodel without the need for additional trainable parameters or mechanisms to reproduce history-dependent behavior. In a related work \cite{Rochaetal2023}, we explore this idea from a different perspective. Instead of learning how to dehomogenize the macroscopic strain, we learn how material properties evolve in time and let the material model be the decoder of a single fictitious material point subjected to the macroscopic strain.

In our previous paper \cite{Maiaetal2023}, the PRNN was demonstrated to work for micromodels with rate-independent plasticity, capturing loading-unloading behavior without seeing it during training. It is anticipated that the same approach can capture rate-dependence. In this paper, we apply the PRNN approach to micromodels where the polymer matrix is described with the Eindhoven Glassy Polymer (EGP) model, an advanced elasto-viscoplastic material model for polymers. For this purpose, the following features are added with respect to the previous work:
\begin{itemize}
\item time-dependent material behavior;
\item a finite strain formulation;
\item generalization to 3D space.
\end{itemize} 

We show how these new and non-trivial features are incorporated into the network and demonstrate that the benefits of the PRNN approach successfully transfer to a much more complex class of models.

\section{Microscale analysis}
\label{sec:micro}

This work focuses on the homogenized behavior of a Representative Volume Element (RVE) of a microscopic material with both path and time-dependency. For notation purposes, the superscripts $\Omega$ and $\omega$ refer to the homogenized (macroscopic) and microscopic quantities, respectively. Let $\omega$ denote the RVE domain and consider that periodic boundary conditions (PBC) are applied to simulate the behaviour of a macroscopic bulk material point. Neglecting body forces, the updated Lagrangian formulation can be defined by the weak statement of equilibrium: 
\begin{equation}
\label{eq:equilibrium} 
\underbrace{\int_{\omega} \mathbf{B}^{\textrm{T}} \boldsymbol{\sigma} \ d \omega}_{\mathbf{f}^{\textrm{int}}} - \underbrace{\int_{\Gamma_\text{u}} \mathbf{N}^{\textrm{T}} \mathbf{t}_\text{p} \ d\Gamma}_{\mathbf{f}^{\textrm{ext}}} = \mathbf{0}
\end{equation}
where $\mathbf{N}$ is a matrix with the shape functions used to interpolate nodal displacements, $\mathbf{B}$ contains the gradients of the shape functions  with respect to the current coordinates $\mathbf{x}$, $\boldsymbol{\sigma}$ is the Cauchy stress, $\mathbf{t}_\text{p}$ are the tractions prescribed on the boundary of the domain $\Gamma_\text{u}$.
\begin{figure}[!h]
\hfill
\subfloat[Boundary conditions of periodic micromodel $\omega$ with constrained dofs to avoid rigid-body motion]{\label{fig:micropbc}
\includegraphics[width=0.375\textwidth]{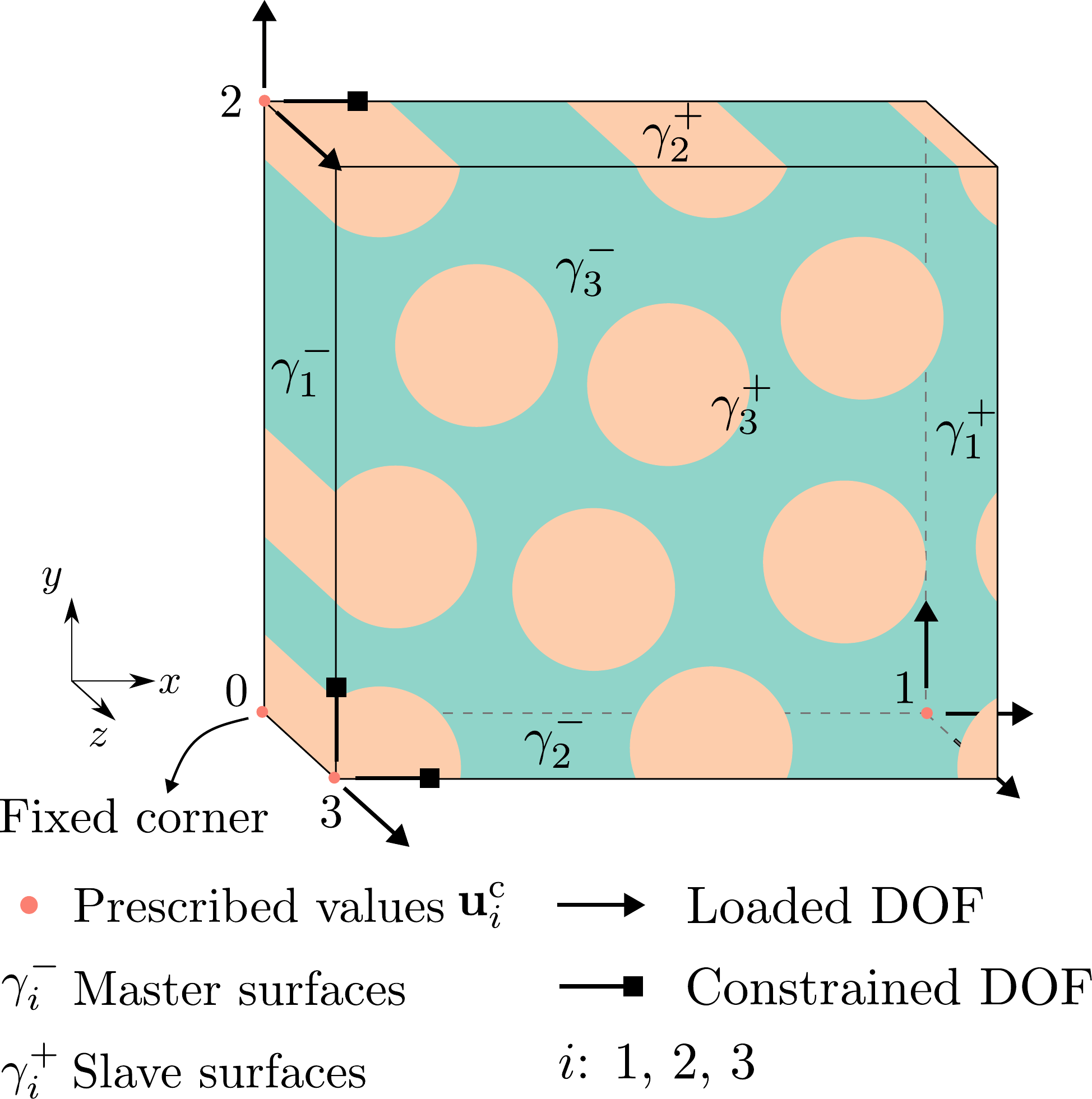}}
\hfill
\subfloat[Initial (undeformed), previously converged (reference) and updated configurations]{\label{fig:updatedlag}
\includegraphics[width=0.55\textwidth]{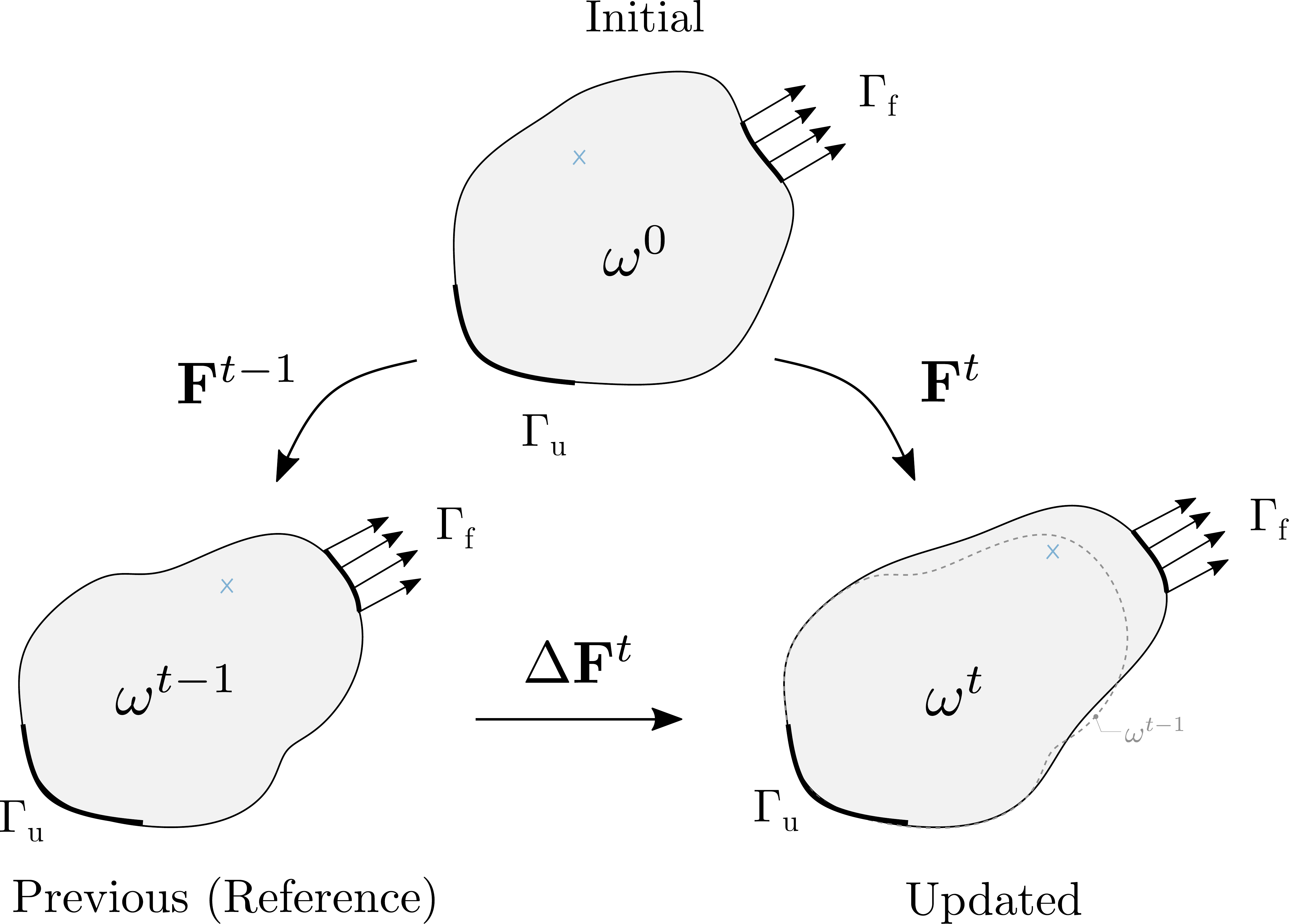}}
\hfill \strut
\caption{Micromodel and scheme of configurations used in the updated Lagrangian framework.}
\label{fig:micromodel}\end{figure}

In the Finite Element (FE) method, \Cref{eq:equilibrium} is solved iteratively as:
\begin{equation}
\label{eq:residual}
\mathbf{r} = \mathbf{f}^{\textrm{int}} - \mathbf{f}^{\textrm{ext}} = \mathbf{0}
\end{equation}  
where $\mathbf{r}$ is the residual vector that vanishes once equilibrium is reached. The iterative procedure involves linearization of $\mathbf{f}^\textrm{int}$ with respect to the degree of freedom vector which in the geometrically nonlinear formulation requires accounting for the dependence of $\mathbf{B}$ from \Cref{eq:equilibrium} on the displacements through a geometric contribution to the stiffness matrix. 

The stress in \Cref{eq:equilibrium} is related to the deformations with a constitutive model $\mathcal{D}^{\Omega}$, which, in general,  can be described by:
\begin{equation}
\label{eq:consteq}
\boldsymbol{\sigma} = \mathcal{D}^{\omega} \ (\mathbf{F}, \intvarvec, \Delta t )
\end{equation}
where $\intvarvec$ are the history variables that account for path and rate-dependency, $\Delta t$ is the time increment and $\mathbf{F}$ is the deformation gradient:
\begin{equation}
\label{eq:defgrad}
\mathbf{F} = \Idtensor + \nabla \mathbf{u}
\end{equation}
where $\nabla \mathbf{u}$ represents the gradient of the microscopic displacements. Since the deformation gradient is calculated with respect to the initial configuration, its increment can also be easily computed from the current and previous deformation states:
\begin{equation}
\label{eq:incdefgrad}
\Delta \mathbf{F} = \mathbf{F} \mathbf{F}^{-1}_{t-1}
\end{equation}
For rate-dependent materials, the stress depends on $\Delta \mathbf{F}$ as well as $\mathbf{F}$, which can be achieved with \Cref{eq:consteq} if $\mathbf{F}_{t-1}$ is included in the material history $\intvarvec$. Upon convergence, the homogenized stresses can be averaged out by integrating the microscopic stresses over the volume $\omega$:
\begin{equation}
\label{eq:stressavtheorem}
\macrostressvec = \frac{1}{\| \, \omega \, \|} \int_{\omega} \boldsymbol{\sigma} \, \text{d}\omega 
\end{equation}

\subsection{Constitutive models}
\label{subsec:constmodels}

In this work, we consider a composite micromodel made of unidirectional fibers embedded in a matrix material. To describe the constitutive behavior of the matrix, the EGP model is used, while for the fibers, a hyperelastic transversely isotropic model is assigned. These consist of the same choices adopted in \cite{Kovacevic2022}, where a thorough validation of the material models was carried out for a carbon/PEEK composite material. Here, we only highlight the main aspects of their formulation and focus on how to incorporate them in a PRNN. 

The fiber constitutive law is based on the one developed by \citet{Bonet1998} with slight modifications \cite{Kovacevic2022}. The constitutive model derives from the strain energy density function and can be split into two components, an isotropic part with a neo-Hookean potential and a transversely isotropic part, with both depending on the right Cauchy-Green deformation tensor:
\begin{equation}
\mathbf{C} = \mathbf{F}^\text{T} \mathbf{F}
\label{eq:rightcauchy}
\end{equation}

The EGP model for the matrix material consists of a rate and path-dependent elasto-viscoplastic, isotropic, 3D constitutive law. In this model, no explicit yield surface is needed since an Eyring-based viscosity function evolves with the stress applied, leading to the viscoplastic flow of the material. The Cauchy stress calculated by the EGP is composed of three contributions: hydrostatic, hardening and driving stress. While the first two parts are defined in more simple terms as they do not depend on the internal variables, in the third part, where viscoplasticity is introduced, a further decomposition can be considered. In this case, the multiple contributions to the driving stress correspond to different molecular (relaxation) processes. Each relaxation process is represented with a series of Maxwell models (modes) connected in parallel, with a shear modulus in the elastic spring and a stress-dependent viscosity in the dashpot. Here, a single relaxation process is considered and represented with 1 mode. 
\begin{figure}[ht!]
\centering
\includegraphics[width=0.5\textwidth]{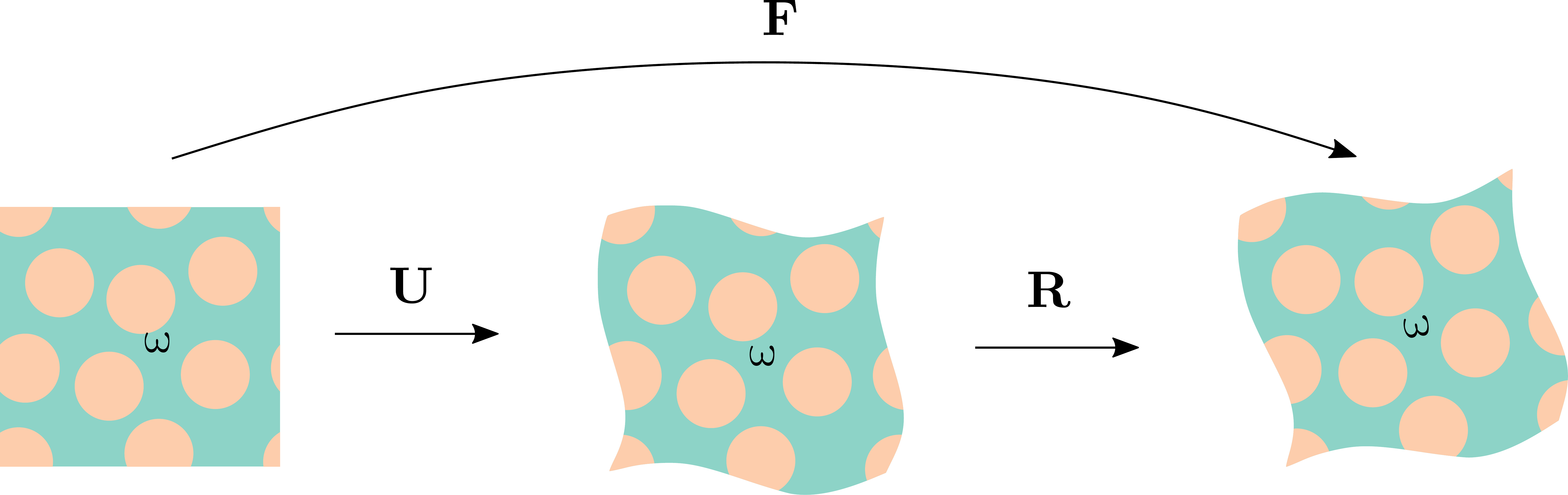}
\caption{Right polar decomposition on deformation gradient $\mathbf{F}$ resulting in the stretch and rotation tensors $\mathbf{U}$ and $\mathbf{R}$, respectively}
\label{fig:polardecomprve}
\end{figure}

For any of the models discussed so far, a helpful tool to deal with the high-dimensionality of the deformation gradient is the polar decomposition theorem. The theorem states that any deformation gradient $\mathbf{F}$ can be uniquely decomposed into the product of two other tensors: a symmetric one $\mathbf{U}$ and an orthogonal one $\mathbf{R}$, as $\mathbf{F}=\mathbf{R}\mathbf{U}$. These two tensors have physical interpretations and are closely related to the principle of material objectivity or material frame indifference. In short, the symmetric tensor represents the deformation (\textit{i.e.} stretches and shear) and the orthogonal tensor represents a rigid body rotation. When applied in this sequence, the final configuration obtained is the same as the one obtained if the deformation gradient was applied directly. 

The particular order of stretch and rotation is known as right polar decomposition and is illustrated in \Cref{fig:polardecomprve}. From these interpretations and considering the principle of material frame indifference, which states the material response is independent of the observer, one can rewrite stresses as:
\begin{equation}
\label{eq:unrotstresses}
\stressunrot = \mathcal{D}^{\omega} \ (\Utensor, \intvarvec, \Delta t)  
\end{equation}
\begin{equation}
\label{eq:rotstresses}
\stressrot = \Rtensor \ \stressunrot \ \Rtensor^\text{T} 
\end{equation}
where $\stressunrot$ are the unrotated stresses and $\stressrot$ are the stresses in the original frame of reference.

\section{Physically Recurrent Neural Network}
\label{sec:prnn}

\begin{figure}
\centering
\includegraphics[width=0.85\textwidth]{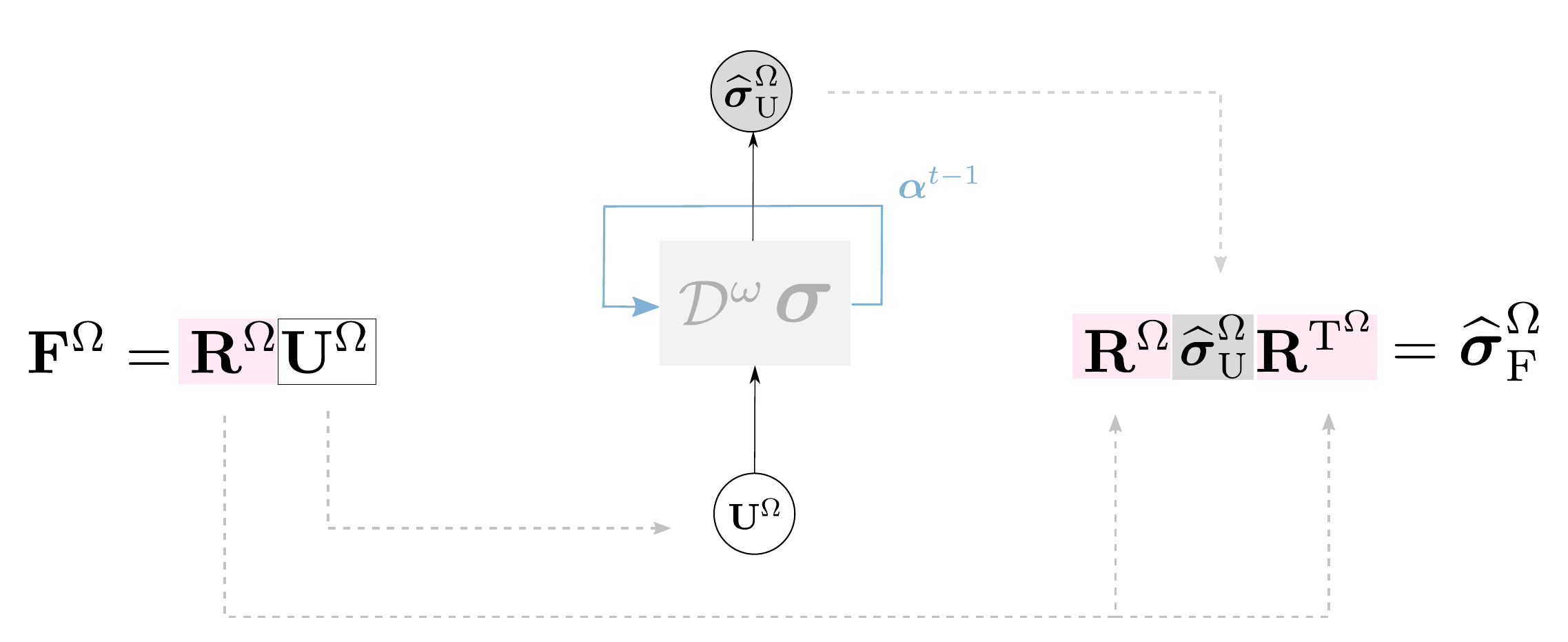}
\caption{Use of PRNN in a general full-order solution setting with $\Ftensor^{\Omega}$ and $\predstressrot$ as input and output, respectively.}
\label{fig:microprnn}
\end{figure}

In this section, we present the new architecture of the Physically Recurrent Neural Network (PRNN) as illustrated in \Cref{fig:prnn}. Having the network in \cite{Maiaetal2023} as the starting point, we extend the model to be used in a 3D finite strain framework considering micromodels with path and rate-dependent behavior. The data-driven parts of the network consist in learning how the homogenized strain can be dehomogenized and distributed among a small set of fictitious material points and how the stress obtained in these material points can be homogenized again. To reduce the dimensionality of the feature space of the network, and consequently to limit the sampling effort required for training, we opt for designing a surrogate for the mapping from stretch (path) to rotated stress (\Cref{eq:unrotstresses}) and let it be embedded between decomposition and rotation operations (see \Cref{fig:microprnn}).

\subsection{Encoder}
\label{subsec:encoder}
\begin{figure}[!h]
\centering
\includegraphics[width = \textwidth]{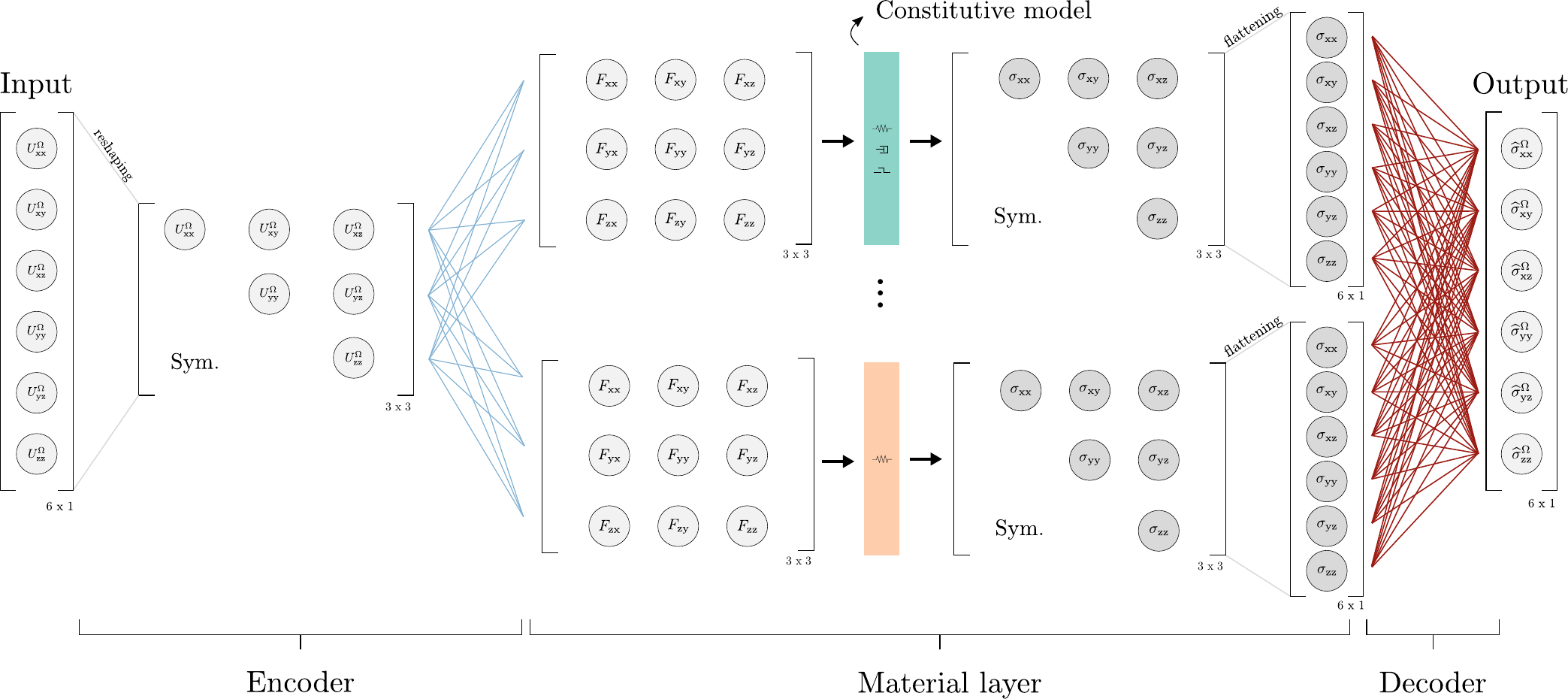}
\caption{New architecture of PRNN for finite strain framework.}
\label{fig:prnn}
\end{figure}

The encoder comprises all parameters and operations that convert the homogenized stretch tensor into local fictitious deformation gradients. In the general PRNN architecture illustrated in \Cref{fig:prnn}, these correspond to the blue connections. In our previous work, the encoder consisted of an arbitrary number of hidden layers fully connected, while in this work a custom layer is proposed to ensure that physical constraints related to the definition of the strain measure are met. Two challenges arise from working with the deformation gradient instead of the small strain vector. Firstly, with the deformation gradient or the stretch, the undeformed state corresponds to the identity and not a null tensor. 

In a regular dense layer, if a given set of weights $\mathbf{W}$ were to be applied on the undeformed stretch tensor (\textit{i.e.} $\Umacrotensor = \Idtensor$), the resulting matrix $\mathbf{W} \Umacrotensor$ would be different from the identity  and therefore generate stresses when it should not. To address that, we need to make a few changes to the encoder, starting with the way we treat the input. Now, instead of applying weights to transform a vector with dimension 6, we work on the actual tensor $\Umacrotensor$ that is $3 \times 3$. Note that this is only a \emph{reshaping} operation, and no additional features are needed to fill the tensor. 

With that, the weights connecting $\Umacrotensor$ to the inputs of the material layer can be applied in a similar fashion to the fictitious material points, in groups, to generate the deformation gradients used in that layer. In this case, for each point, a $3 \times 3$ weight matrix is needed. Another important change to ensure the zero stress-state comes from the definition of the deformation gradient (see \Cref{eq:defgrad}). Based on that, we subtract the identity matrix from the homogenized stretch tensor and only then apply the weights to the remaining values. After that transformation, we add the identity back and obtain the final deformation gradient.

\begin{figure}[!t]
\centering
\includegraphics[width=\textwidth]{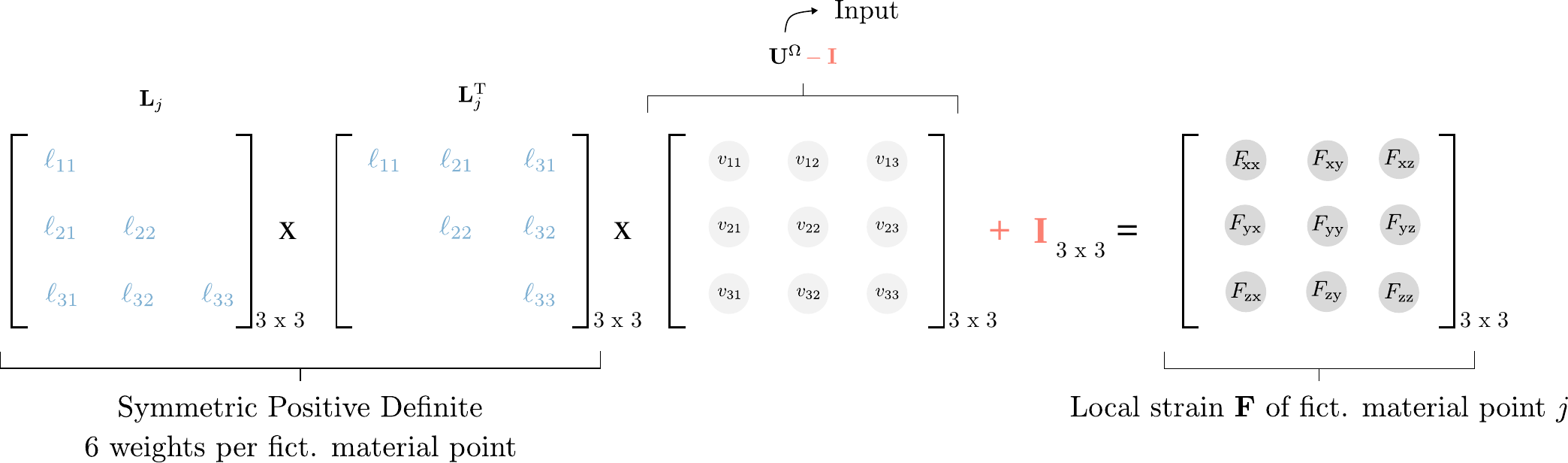}
\caption{Encoder architecture applied to obtain the local strain of a fictitious material point $j$ based on the input $\mathbf{U}^{\Omega}$.}
\label{fig:encoder}
\end{figure}

Secondly, because the determinant represents the change in volume from the undeformed to the current configuration, the local deformation gradients learned by the network should have strictly positive determinants. One way to avoid negative determinants consists in ensuring that the determinant of the weight matrices applied on $\Umacrotensor-\Idtensor$ to obtain the fictitious local deformation gradients are always positive. This is done by imposing a structured weight matrix originated from a Cholesky decomposition for each subgroup $j$:
\begin{equation}
\label{eq:ludecomp}
\Wjmat = \mathbf{L}_{j}\mathbf{L}^\text{T}_{j} = \begin{bmatrix} \ell_{11} &  & \\ \ell_{21} & \ell_{22} & \\  \ell_{31} & \ell_{32} & \ell_{33} \end{bmatrix} 
\begin{bmatrix} \ell_{11} & \ell_{21} & \ell_{31} \\ & \ell_{22} & \ell_{32} \\ & & \ell_{33} \end{bmatrix} 
\end{equation}
The determinant of the decomposed triangular matrices simplifies to the multiplication of their diagonal elements, so positivity is therefore ensured by applying a softplus function to those diagonal entries. In this case, only 6 learnable parameters are associated to each fictitious material point. The scheme in \Cref{fig:encoder} summarizes how the local strain of one fictitious material point is obtained after the proposed changes. 

\subsection{Material layer}
\label{subsec:matlayer}
This layer contains the embedded physics-based constitutive models, arranged into a series of fictitious integration (material) points. Because a material model is not a scalar-valued function like typical activation functions (\textit{e.g.} sigmoid, tanh, relu, etc.), a special architecture is required. In that sense, an important change compared to \cite{Maiaetal2023} is the way neurons are interpreted. Here, we group them together in $m$ subgroups, each consisting of a tensor with the same order tensor and dimensions as the deformation gradient in the input layer (3 $\times$ 3 for the present investigation in three dimensions), whereas in \cite{Maiaetal2023} the subgroups consists of vectors of length 3, representing the strain vector in 2D. 
In this arrangement, each subgroup corresponds to one \textit{fictitious material point}. The basic idea is that the values reaching the subgroup can be seen as a local deformation learned by the encoder, which will then be evaluated by one of the constitutive models used in the full-order solution with the same material properties. 

Once a given constitutive model with its respective material properties is assigned to the subgroup $j$, say $\mathcal{D}^{\omega}_j$, the next step is to use it to obtain the stresses and the updated internal variables (if any). These internal variables are present in rate and path-dependent material models and are the core of the physics-based memory of the proposed network. However, rate and path-independent constitutive models can also be used in the material layer without further adaptations. A brief discussion on the choice of the constitutive models used in this layer follows at the end of the section.

\begin{figure}[!t]
\centering
\includegraphics[width=.5\textwidth]{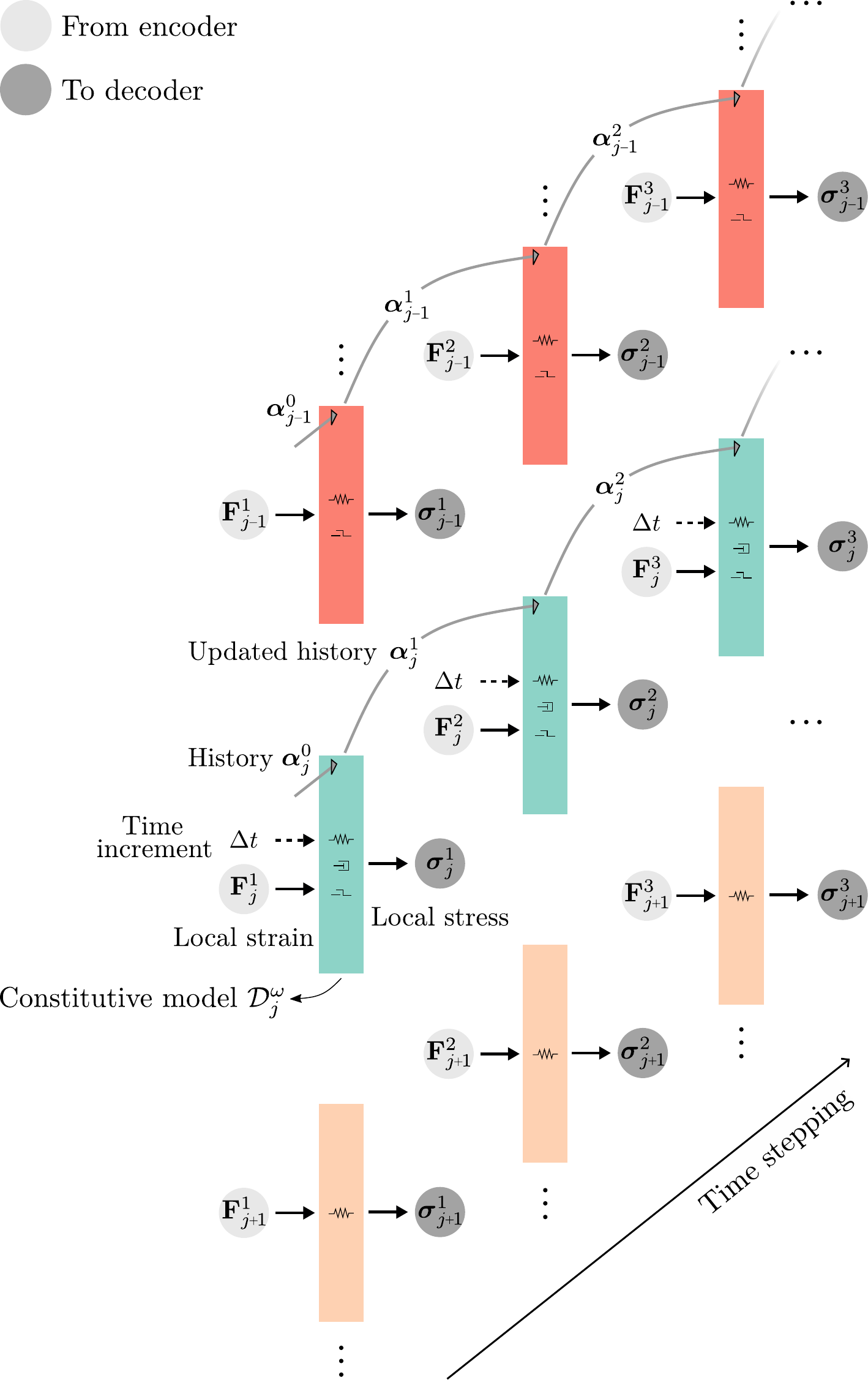}
\caption{Scheme of fictitious material points unrolled in time, each colored box corresponding to a different constitutive model. From top to bottom: path-dependent, path and rate-dependent, and path and rate-independent constitutive models.}
\label{fig:newprnncell}
\end{figure}

Consider that $\mathcal{D}^{\omega}_j$ takes as input the deformation gradient $\mathbf{F}$, the internal variables from previous time step $\intvarvec^{t-1}$ and the increment of time $\Delta t$. In the first time step, the internal variables of all material points are properly initialized based on the undeformed state $\boldsymbol{\alpha}_j^0$. In every time step, the current stresses $\boldsymbol{\sigma}$ and updated internal variables $\intvarvec$ of each subgroup are obtained. These variables are preserved in each subgroup so that in the following load step, when a new $\mathbf{F}$ is fed to the material point, the history of the material can be updated accordingly. A representation of this workflow is shown in \Cref{fig:newprnncell}. Note that the ``flattening" operation transforming the $3 \times 3$ tensor into a vector with only the 6 independent components, is analogous to the reshaping operation used at the encoder. This condensation does not imply in loss of information since the Cauchy stress tensor is symmetric.

An important aspect illustrated in \Cref{fig:newprnncell} is that no additional time-related features or trainable parameters are needed to learn the time-dependence. The network learns the strain distribution over the fictitious material points through the encoder, which works the same for all constitutive models. The time increment $\Delta t$ is passed to the rate-dependent material as additional input, but it has the same value for all material points as would be done in the micromodel simulation. By directly employing the same material models and properties considered in the micromodel with internal variables that naturally follow physics-based assumptions, we can capture the rate and path-dependent behavior in a more straightforward way. With RNNs, the mechanisms behind the evolution of internal variables need to be learnt from the data.

Finally, the user is left with the choice of which constitutive model to employ in the material points. Our recommendation is to employ all nonlinear constitutive models used in the micromodel with their respective known material properties. To illustrate that, consider the composite micromodel studied in the numerical examples of this work, in which an orthotropic hyperelastic model is used to describe the fibers and an elasto-viscoplastic model for the matrix. Since both models include nonlinearity in their formulations, we include both types in the material layer. In addition to that, in the present case, each model has distinctive behavior in terms of path and rate-dependence, which emphasizes the importance of both in the network. This topic is further discussed in \Cref{subsec:training} along with other training aspects and model selection procedure, including the definition of the proportion of the constitutive models in this layer.  

\subsection{Decoder}
\label{subsec:decoder}
The decoder comprises all network parameters that work on the outputs of the material layer to obtain the homogenized stresses $\widehat{\boldsymbol{\sigma}}^{\Omega}$. Note that because the outputs of the material layer consist of the stresses from the fictitious material points, the role of the decoder parameters is well aligned with the actual full-order solution. In the micromodel, once the full-field of stresses is obtained, the homogenized stresses are obtained by averaging the stresses over the entire domain. Here, instead of integrating the field with hundreds or thousands of integration points, only a few fictitious material points contribute to the homogenized response where the relative contributions of each fictitious point are learnt from data.

For that purpose, an arbitrary number of neurons and layers can be used. In this work in particular, for a more direct analogy with the homogenization process, a single dense layer with linear activation and physics-motivated modifications is considered. In this way, the weights of the output layer reflect the relative contribution of each of the material points to the predicted homogenized response. In the actual micromodel, weights come from a numerical integration scheme and are strictly positive. To reflect that, an absolute function $\rho \, (\cdot)$ is applied element-wise on the weights of the decoder $\mathbf{W}_\text{d}$. For the present architecture (see \Cref{fig:prnn}), it then follows that the predicted homogenized stress is given by:
\begin{equation}
\label{eq:activateweights}
\widehat{\boldsymbol{\sigma}}^{\Omega} = \rho \, (\weights_\text{d}) \ \mathbf{a}
\end{equation}
where $\mathbf{a}$ corresponds to the concatenation of local stresses from the material points. 
\begin{figure}[!h]
\centering
\includegraphics[width=\textwidth]{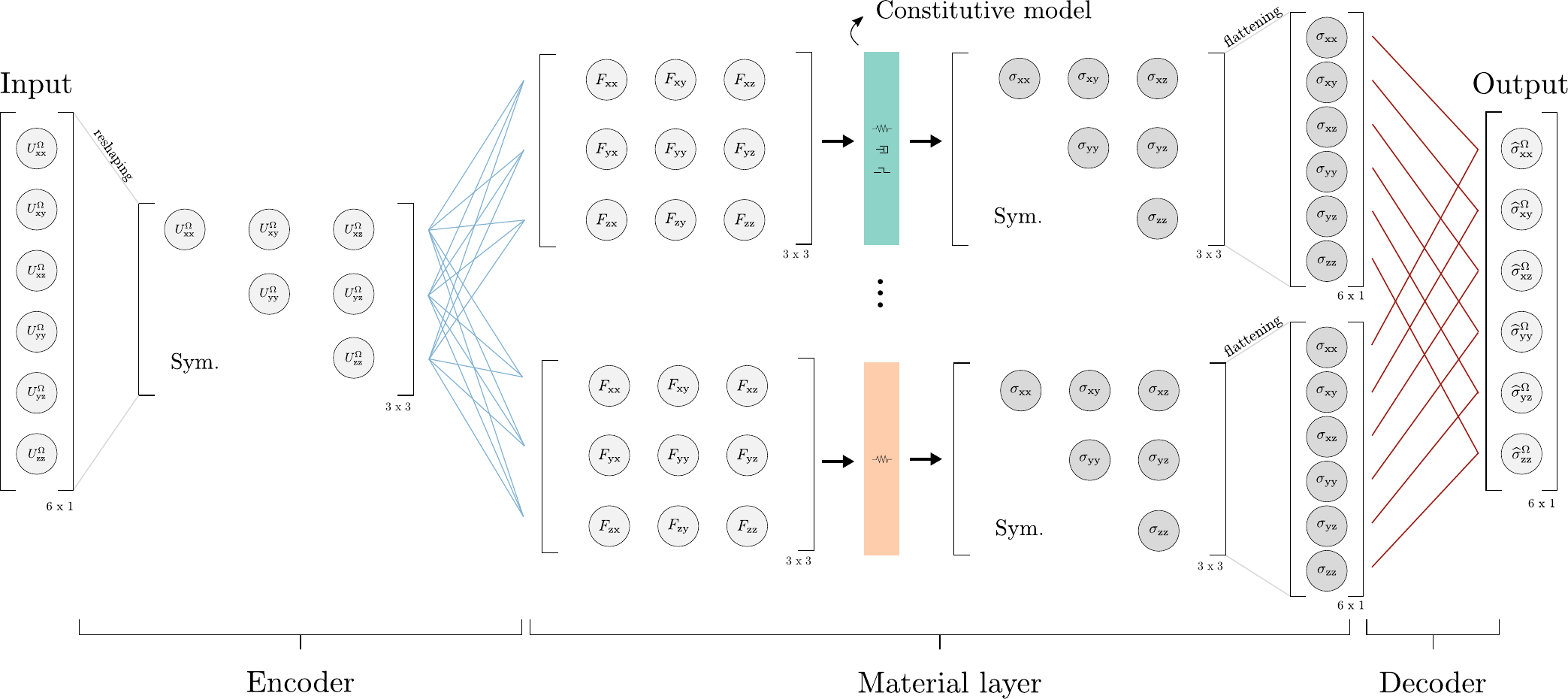}
\caption{PRNN with sparse decoder.}
\label{fig:sparsedecoder}
\end{figure}

In addition to that, we also investigate the use of a sparsification approach, where instead of having a regular dense layer that connects all components of the local stress tensor to the predicted homogenized stress, only the component-wise contributions are taken into account, as illustrated in \Cref{fig:sparsedecoder}. For instance, only the stresses $\sigma_\text{xy}$ from each of the subgroups are weighted in for obtaining $\widehat{\sigma}_\text{xy}^{\Omega}$. This sparsification also brings the decoder closer to the actual homogenization procedure, in which stresses are averaged component-wise.

\subsection{Training aspects and error metrics}
\label{subsec:training}
The goal of the optimization procedure is to minimize a loss function that quantifies how close the network's prediction are from the actual solution. In this work, the standard loss function based on the  mean square error (MSE) is used:
\begin{equation}
\label{eq:loss}
L = \frac{1}{N_\text{train}} \sum_{t=1}^{N_\text{train}} \frac{1}{2} \norm{\, \text{vec} \, ( \boldsymbol{\sigma}^{\Omega}_t) - \text{vec} \left( \widehat{\boldsymbol{\sigma}}^{\Omega}_t \left( \mathbf{U}_t^{\Omega}, \mathbf{W}, \mathbf{W}_\text{d} \right) \, \right) \,}^2
\end{equation}
where $N_\text{train}$ is the number of loading paths used for training, $\mathbf{W}$ and $\mathbf{W}_\text{d}$ are the network parameters for the encoder and decoder, respectively, and vec($\cdot$) corresponds to the Voigt representation of the homogenized stress tensor, which consists of 6 components in the 3D case (\textit{i.e.} the ``flattening" mentioned in the previous sections). 
From that, one can compute the gradients of the loss function with respect to the trainable parameters using a backpropagation procedure and then update those accordingly, for which we use the Adam optimizer \cite{Adametal2014}. 

The backpropagation here follows the same methodology as in \cite{Maiaetal2023}. Note that the gradients of the parameters in the decoder can be obtained based on the conventional backpropagation procedure, but for the ones in the encoder, backpropagation through time is needed. This is a vital aspect of the training and stems from the path-dependency of the material models embedded in the material layer. For completeness, we include the expression for computing the gradients of the weights in the encoder at time step $t$ for a given loading path:  
\begin{equation}
\label{eq:derivenc}
\frac{\partial L^t}{\partial \Ljmat}=\pd{L}{\predmacrostress} \pd{\predmacrostress}{\stress[j]{t}} \left\{\pd{\stress[j]{t}}{\defgrad[j]{t}} \pd{\defgrad[j]{t}}{\Ljmat}+\pd{\stress[j]{t}}{\intvar[j]{t}} \pd{\intvar[j]{t}}{\defgrad[j]{t}} \pd{\defgrad[j]{t}}{\Ljmat}+\pd{\stress[j]{t}}{\intvar[j]{t}} \sum_{\bar{t}=1}^{t-1}\left[\left(\prod_{\tilde{t}=\bar{t}+1}^{t} \pd{\intvar[j]{\tilde{t}}}{\intvar[j]{\tilde{t}-1}}\right) \pd{\intvar[j]{\bar{t}}}{\defgrad[j]{\bar{t}}} \pd{\defgrad[j]{\bar{t}}}{\Ljmat}\right]\right\}
\end{equation}
where $\Ljmat$ corresponds to the weights associated to the material point $j$. The gradients related to the internal variables are evaluated using central finite differences. However, if the material models are implemented with automatic differentation support (\textit{e.g.} PyTorch and TensorFlow), these gradients and dependencies can be automatically taken into account with tools such as Autograd and GradientTape, as done with off-the-shelf RNNs. 

A potential issue in training with \Cref{eq:loss} is the large variations of values across the multiple outputs due to the orthotropy of the composite material with high stiffness contrast. In such scenario, one component can disproportionately dominate over the others, leading to unstabilities in the training process and overall poor performance. To mitigate that, each component of $\boldsymbol{\sigma}^{\Omega}$ is normalized to [-1, 1] as follows: 
\begin{equation}
\label{eq:normalization}
\sigma^{\Omega}_{(\cdot) \, \text{norm}} = 2 \, \left( \cfrac{\sigma^{\Omega}_{(\cdot)} - \text{min} \, \sigma^{\Omega}_{(\cdot)}}{\text{max} \,  \sigma^{\Omega}_{(\cdot)} - \text{min} \, \sigma^{\Omega}_{(\cdot)}} \right) - 1
\end{equation}
where max refers to the maximum absolute homogenized stress values of the component $(\cdot)$ in the training data and min is the negative of that value. The symmetric bounds in each of the components ensures that the zero-stress state from the material points will be reflected in the homogenized stress. Furthermore, to preserve the role of the decoder as the homogenization-like step, the normalization in \Cref{eq:normalization} is also applied to the local stresses from the fictitious material points. This ensures that all material point stresses are within the same range expected at the output layer.  Lastly, no normalization is considered for the inputs, since the range of the features are similar and, more importantly, are compatible with the range expected by the models in the material layer.

For the model selection and performance assessment, we consider two error metrics:
\begin{equation}
\label{eq:error}
\begin{aligned}
\text{Absolute error}: \frac{1}{N_\text{paths}}\sum_{i=1}^{N_\text{paths}} \frac{1}{L_\text{path}} \sum_{t=1}^{L_\text{path}} \norm{\, \text{vec} \, ( \boldsymbol{\sigma}^{\Omega}_t) - \text{vec} \, ( \widehat{\boldsymbol{\sigma}}^{\Omega}_t ) \,} \\ 
\text{Relative error}: \frac{1}{N_\text{paths}}\sum_{i=1}^{N_\text{paths}} \frac{1}{L_\text{path}} \sum_{t=1}^{L_\text{path}} \frac{\norm{\, \text{vec} \, ( \boldsymbol{\sigma}^{\Omega}_t) - \text{vec} \, ( \widehat{\boldsymbol{\sigma}}^{\Omega}_t) \,}} {\norm{\, \text{vec} \, ( \boldsymbol{\sigma}^{\Omega}_t) + \boldsymbol{\varepsilon} \,} }
\end{aligned}
\end{equation}
where $N_\text{paths}$ refers to the number of loading paths in the validation/test sets, $L_\text{path}$ is the length of each path and $\boldsymbol{\varepsilon}$ is a stabilizing term with the same dimensions as $\text{vec} \, (\boldsymbol{\sigma}^{\Omega}_t)$ filled with $10^{-8}$ used to avoid division by zero. 

To reduce the number of hyper-parameters to be tuned and keep the model selection as simple and straightforward as possible, we define a minibatch as 2 paths, the stopping criterion as the maximum number of epochs (\qty{1000}) and use the recommended parameters from \cite{Adametal2014} in the Adam optimizer. When training the network, the validation set is evaluated every 50 epochs, and the best set of parameters is updated only if the current error is lower than the historical lowest validation error, thus mitigating the risk of overfitting. Further details on the model selection procedure, including the definition of the material layer size, the type of decoder (dense or sparse), and the size of the training set, are presented in \Cref{sec:numexp}.

Finally, when choosing which constitutive models to assign to the fictitious material points, we follow the idea of including all sources of non-linearity, which in this case consists of both the fiber and the matrix constitutive models. At this point, it is worth highlighting another aspect that makes having both models in the network important. Although the fiber constitutive model adopted in this work only shows non-linearity at very large strains, in our case, it is also the one introducing the transversal isotropy in the micromodel and has distinct behavior from the matrix in terms of path and rate-dependency. Those unique characteristics need to be present in the network so that the encoder and decoder can leverage them into the homogenized stress response. Related to that is the definition of how many of the fictitious material points are assigned to each of the models. This proportion itself is a hyper-parameter, but to reduce the amount of variables in the upcoming studies, we define a fixed splitting ratio.  The hyperelastic and elasto-viscoplastic models correspond to \qty{25}{\%} and \qty{75}{\%} of the material points, respectively, rounding the number of hyperelastic models up when the total number of points is even but not divisible by 4.

\subsection{Use as constitutive model}
\label{subsec:constmodel}

For incorporating the present network as a constitutive model in a microscale analysis that takes as input the homogenized deformation gradient ($\Fmacrotensor$) and the increment of time ($\Delta t$) and outputs homogenized stresses $\predstressrot$, a few additional steps are introduced. First, the polar decomposition theorem is applied on the deformation gradient in order to obtain the rotation $\Rmacrotensor$ and the stretch tensors $\Umacrotensor$. Once the stretch tensor is obtained and the increment of time is known, the network is used to predicted the unrotated stresses $\widehat{\boldsymbol{\sigma}}^{\Omega}_\textrm{U}$ in a forward pass. The rotation tensor is then used to transform the predicted unrotated stresses back into the rotated system.

Obtaining the tangent stiffness matrix is not as straightforward. In this framework, the jacobian of the network is only one part of the tangent stiffness matrix expression for the entire mapping between rotated stresses and the deformation gradient:  
\begin{equation}
\label{eq:tangentmicroprnn}
\pd{\predstressrot}{\Fmacrotensor} = \pd{\predstressrot}{\predstressunrot} \ \pd{\predstressunrot}{\Umacrotensor} \ \pd{\Umacrotensor}{\Fmacrotensor} + \pd{\predstressrot}{\Rmacrotensor} \ \pd{\Rmacrotensor}{\Fmacrotensor}
\end{equation}
where the partial derivatives of the homogenized rotation and stretch tensors with respect to the homogenized deformation gradient are given by the expressions derived by \citet{Chen1993} and $\partial{\predstressunrot}/\partial{\Utensor}^{\Omega}$ is given by performing a complete backward pass through the network. Moreover, the partial derivative of the stresses with respect to the unrotated stresses is given by:
\begin{equation}
\label{eq:dstressrotdstress}
\pd{\predstressrot}{\predstressunrot} = \Rmacrotensor \otimes \Rmacrotensor \\
\end{equation}
where $\otimes$ represents the Kronecker product between two second-order tensors of dimensions $n_\text{rank} \times n_\text{rank}$, resulting in a second-order tensor of dimensions $n_\text{rank} \ n_\text{rank} \times n_\text{rank} \ n_\text{rank}$. Finally, the partial derivative of the stresses with respect to the rotation tensor are evaluated as:
\begin{equation}
\label{eq:dstressdr}
\pd{\predstressrot}{\Rmacrotensor} = \bar{\mathbf{P}} \ ( \Idtensor \otimes \predstressunrot \ \mathbf{R}^{\Omega^\textrm{T}}) + (\Idtensor \otimes \Rmacrotensor \ \predstressunrot ) \ \mathbf{P}
\end{equation}
where $\bar{\mathbf{P}}$ and $\mathbf{P}$ are two permutation matrices given by:
\begin{equation}
\label{eq:permmatrices}
\begin{aligned}
\bar{\mathbf{P}} = \sum_{i,j} E_{ij} \otimes E_{ij} \\
\mathbf{P} = \sum_{i,j} E_{ij} \otimes E_{ji}
\end{aligned} 
\end{equation}
with $E_{ij}$ being a null matrix except for the unit value at $E_{i,j}$. 

\section{Data generation}
\label{sec:datagen}

In general, surrogate models need to be trained with an extensive amount of data covering several types of loading. This is because it is virtually impossible to have fine control over what types of loading the micromodel will experience upfront even in the simplest scenarios. Therefore, to investigate how well the proposed network can generalize to unseen scenarios, a variety of loading functions and methods for generating the loading paths are considered. 

\begin{figure}[!h]
\centering
\includegraphics[width=.23\textwidth]{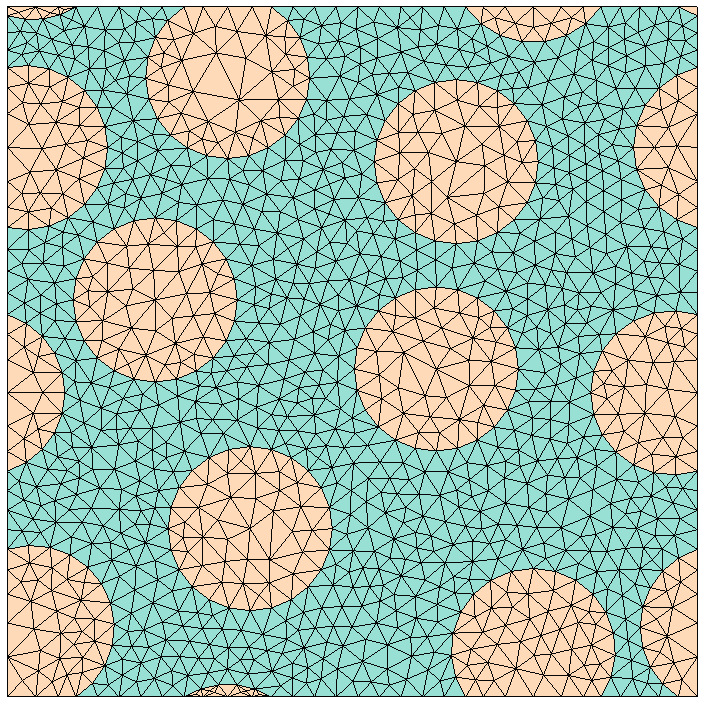}
\caption{Geometry and mesh discretization of micromodel used to generate the data.}
\label{fig:rvemesh}
\end{figure}

First, we define the geometry and the discretization of the micromodel. In this case, the same composite RVE used in \cite{Kovacevic2022}, and illustrated in \Cref{fig:rvemesh}, with 9 fibers embedded in a matrix material is adopted. The material models and properties assigned to each of the phases also follow from that work with a minor change in one of the material properties of the matrix. The reinforcements are assumed to be carbon fibers and can be described by the hyperelastic, transversely isotropic material model developed by \cite{Bonet1998}. For the matrix, the elasto-viscoplastic EGP model is considered with the relaxation spectrum now consisting of one mode (the first). Both of these models are briefly discussed  in \Cref{subsec:constmodels}, but for further details on their implementation and numerical validation in the 3D finite strain framework, the reader is directed to the reference paper \cite{Kovacevic2022}. 

To generate the data, two strategies are devised, one producing proportional loading paths, and the other non-proportional loading paths. We use the first type to train and test the network, while the second is reserved for testing only. By proportional we refer to curves in which the loading direction is fixed. For this, we adopt the arc-length formulation with indirect displacement control derived in \cite{Rocha2020}, in which a constant unit load vector is considered and the additional constraint consists in the unsigned sum of the controlled displacements. For stress measures based on the undeformed state, that also entails a constant stress ratio. 

For creating proportional paths, three main ingredients are needed: the loading direction $\mathbf{n}$, the loading function $\lambda$ and the time increment $\Delta t$. In the previous work \cite{Maiaetal2023}, basic load cases (\textit{e.g.} uniaxial and biaxial tension and compression, transverse and longitudinal shear, etc.) were used for training PRNNs subjected to general stress states. Here, due to the increased problem dimensionality, we train with a more general approach of random loading directions. For each path, the unit load vector is obtained by sampling values from 6 independent Gaussian distributions (X $\sim \mathcal{N}(0, 1)$) and normalizing them to a unit vector, one for each prescribed corner displacement. As for the time increment, we set it to $\Delta t = \qty{1}{\s}$ for all time steps. Fixing the time increments allows for a straigthforward assessment of the ability of the network to extrapolate to unseen strain-rates.  
\begin{figure}[!h]
\centering
\includegraphics[width=.85\textwidth]{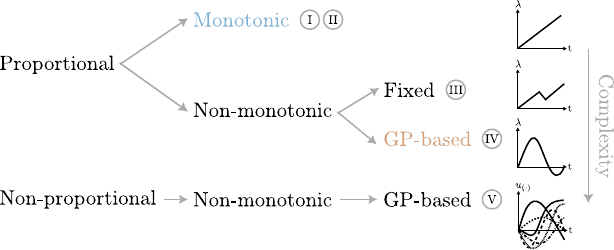}
\caption{Scheme of loading types considered in this work, with colored types being used for training and testing, while remaining are for testing.}
\label{fig:loadtypes}
\end{figure}

\begin{figure}[!ht]
\centering
\subfloat[Monotonic loading (type I)]{\label{fig:doetype1}
\includegraphics[width=0.33\textwidth]{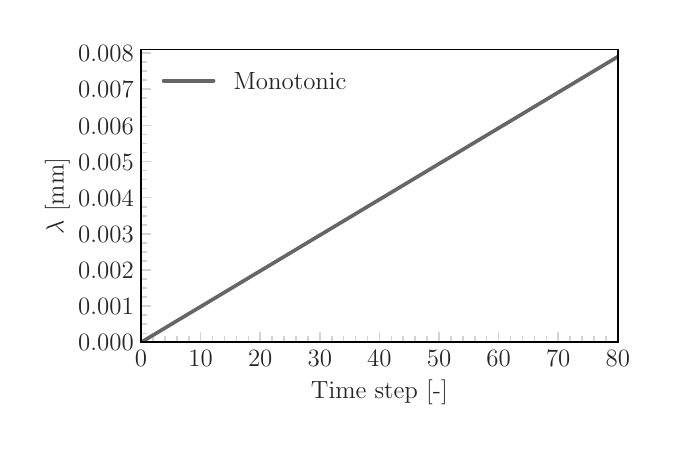}}
\subfloat[Non-monotonic loading (type III)]{\label{fig:doetype2}
\includegraphics[width=0.33\textwidth]{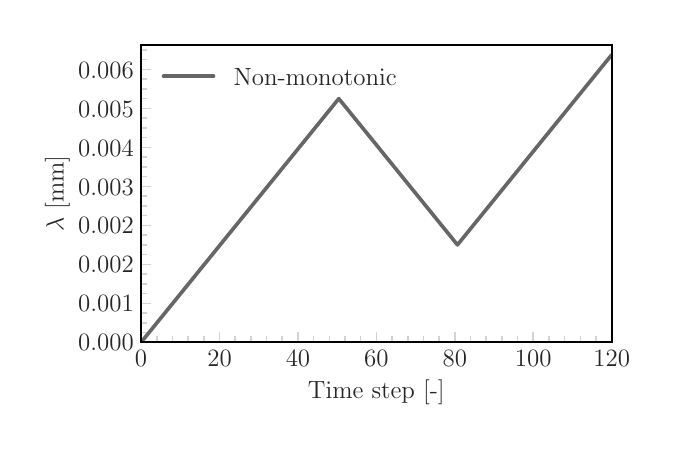}}
\subfloat[Proportional GP-based curves (type IV)]{\label{fig:doetype5}
\includegraphics[width=0.33\textwidth]{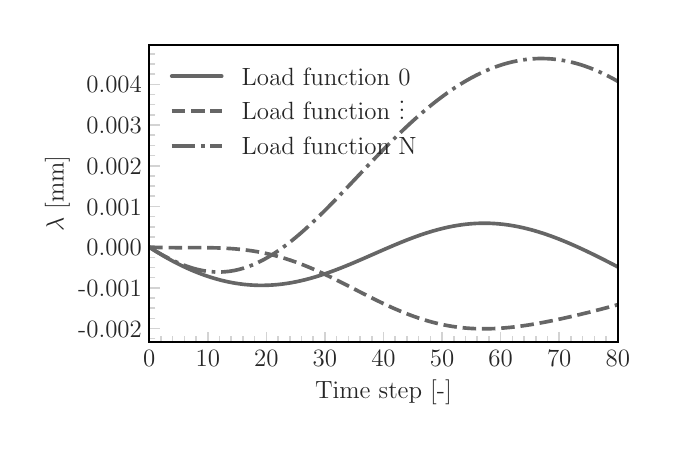}}
\caption{Loading functions used to create proportional loading paths.}
\label{fig:shapefunctions}
\end{figure}

The last ingredient to create the proportional curves is the loading function $\lambda$. We use the two loading functions depicted in \Cref{fig:doetype1,fig:doetype2} as pre-defined monotonic and non-monotonic curves, respectively. Although useful for testing, this non-monotonic set is not as valuable for training since all curves follow the same unloading/reloading behavior. An  alternative with more unloading variety is to sample $\lambda$ from a Gaussian Process (GP) with X $\sim \mathcal{N}(\mu,\,\sigma^{2})$ and covariance function given by: 
\begin{equation}
\label{eq:gpcurves}
k(\xvec_p, \xvec_q) = \sigma_f^2 \exp \Big( - \frac{1}{2 \ell^2} \|  \xvec_p - \xvec_q \|^2 \Big)
\end{equation}
where $\xvec_p$ and $\xvec_q$ are the time step indices of the sequence of loading function values, $\sigma_f^2$ is the variance and $\ell$ is the length scale. These hyper-parameters control the smoothness and how large the unsign sum of the controlled displacements can be, and are tuned to obtain smooth loading functions, as the ones illustrated in \Cref{fig:doetype5}. 

To create a more diverse set in terms of strain-rate compared to the curves using a single pre-defined loading function, for the proportional GP-based curves, the time increment of each path is drawn from a bounded uniform distribution $\Delta t \sim U \, (\qty{0.01}{\s}, \qty{100}{\s})$. \Cref{fig:loadtypes} shows a summary of the three loading types discussed so far ordered by their level of complexity. In this work, we train with two of them, namely  monotonic curves (in blue) and proportional GP-based curves (in brown). For testing, we take a step further and generate non-proportional and non-monotonic paths. These are the most complex paths considered and also employ GPs in their formulation. To create these curves, first we switch to a displacement control method and follow a similar procedure as the one employed in \cite{Maiaetal2023}. Here, we sample the displacements at the controlling nodes from 6 independent GPs, allowing unloading/reloading to take place at different times across the components of the homogenized deformation gradient. This is illustrated in the bottom right plot of \Cref{fig:loadtypes}, where independent $u_{(\cdot)}$--$t$ functions are plotted for the different components. 

For reference, all the types of loading paths studied in the following section are listed below in ascending order of complexity:
\begin{itemize}
\item Type I: proportional and monotonic loading path. The direction $\mathbf{n}$ is generated randomly, the loading function $\lambda$ is as illustrated in \Cref{fig:doetype1} with step size $\Delta \lambda = \qty{1e-4}{\mm}$, and $\Delta t = \qty{1}{\s}$. In the following sections, data sets using this type of path carry the subscript ``mono";
\item Type II: proportional and monotonic loading path with same loading function and step size as Type I, but different strain-rate. Data sets with this type of path carry the subscript ``mono" and two variations of superscript, ``faster" and ``slower". To generate those, $\Delta t = \qty{0.01}{\s} $ and $\Delta t = \qty{100}{\s}$ are used, respectively;
\item Type III: proportional and non-monotonic loading path with fixed unloading/reloading behavior $\lambda$ as illustrated in Fig. \ref{fig:doetype2} with $\Delta \lambda = \qty{1e-4}{\mm}$, and $\Delta t = \qty{1}{\s}$. Data sets with this type of path have the subscript ``unl" and the superscript ``fixed";
\item Type IV: proportional and non-monotonic loading path with loading function given by a GP with variable step size, and $\Delta t\sim U \, (\qty{0.01}{\s}, \qty{100}{\s})$. In this case, each loading path follows a different unloading/reloading function. Fig. \ref{fig:doetype5} illustrates some of the loading functions generated by this approach with $\ell = 30$ and $\sigma_f^2 = \qty{1e-5}{}$ as the hyper-parameters of the GP. Data sets with this type of path have the subscript ``unl" and the superscript ``prop. GP";
\item Type V: non-proportional and non-monotonic loading path with GPs to describe the displacements, and $\Delta t\sim U \, (\qty{0.01}{\s}, \qty{100}{\s})$. Each controlled displacement in the micromodel is assigned to an independent GP, from which we sample smooth and random functions with variable step size. In this case, the hyper-parameters of the GPs are $\ell = 30$ and $\sigma_f^2 = 2.5 \cdot 10^{-7}$, with the exception of the variance of the GP associated to the displacement in the fiber direction, which is 10 times smaller than the others to prevent excessively high stress values that can dominate the homogenized stress state. Data sets with this type of path have the subscript ``unl" and the superscript ``non-prop GP".
\end{itemize}

\section{Numerical experiments}
\label{sec:numexp}
In this section, the accuracy of the network is assessed in a set of numerical experiments. The goal is to illustrate the extrapolation properties of the method given the different training strategies. The test cases cover loading directions and strain-rates different from those seen in training, as well as complex unloading/reloading cases. Since we are focusing on the network's accuracy only, the following sections deal with the stretch and the unrotated stresses as their inputs and outputs, respectively.
\begin{figure}[!h]
\centering
\includegraphics[width=0.7\textwidth]{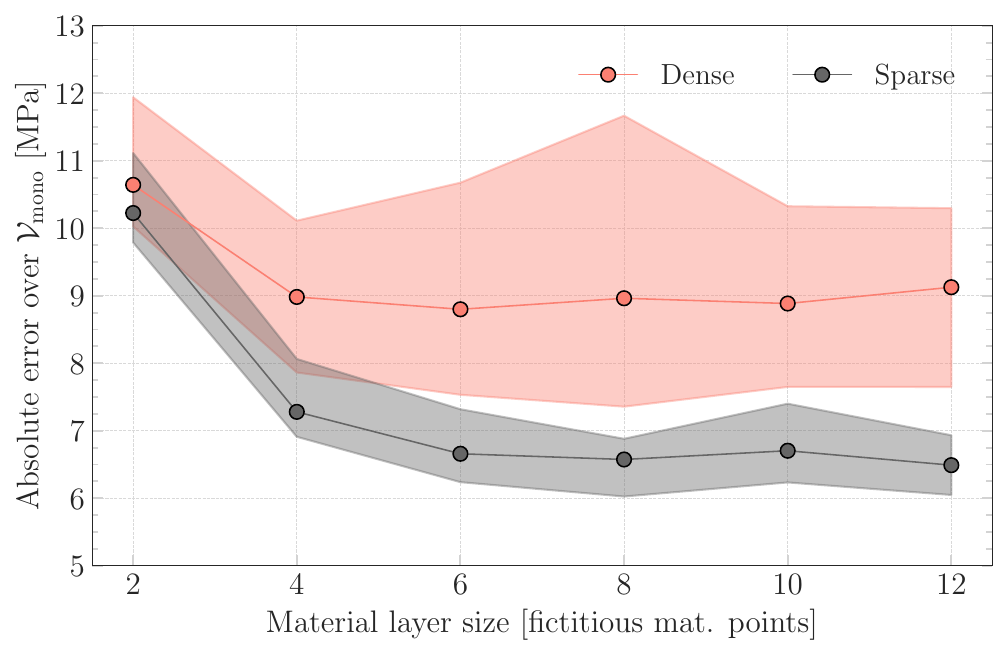}
\caption{Envelope of highest and lowest absolute validation error from 10 PRNNs trained on $\mathcal{D}_{\textrm{mono}} = \{$144 monotonic curves$\}$ over validation set $\mathcal{V}_{\textrm{mono}} = \{$100 monotonic curves$\}$ with different material layer sizes and decoder architectures.}
\label{fig:decoders}
\end{figure}

\subsection{Model selection}
\label{subsec:modelsel}
  
First, two preliminary studies are carried out for model selection. The first one is used to choose between sparse and dense decoders (see \Cref{subsec:decoder}), while the second is focused on defining the material layer size. The comparison is carried out with varying size of the material layer each time considering the largest training set with monotonic loading paths. For each combination of decoder architecture and material layer size, 10 random initializations of the PRNN are considered. In each of them, the training set $\mathcal{D}_{\textrm{mono}}$ consists of 100 monotonic curves randomly selected from a pool of  1000 curves of the same type (Type I). For validation, a fixed set $\mathcal{V}_{\textrm{mono}}$ with  100 monotonic curves is considered. In Fig. \ref{fig:decoders}, the colored areas correspond to the envelope with the highest and lowest absolute errors for each combination, along with the average errors represented by the solid lines with markers. In this case, a marked difference in accuracy between the two types of decoder for all range of material layer size over $\mathcal{V}_{\textrm{mono}}$ is observed. Therefore, in the remainder of this paper, all networks have a sparse decoder. 

\begin{figure}[!ht]
\hfill
\subfloat[$\mathcal{V}_{\textrm{mono}} =\{$100 monotonic loading paths$\}$]{\label{fig:valmono} \includegraphics[height = 5.8cm]{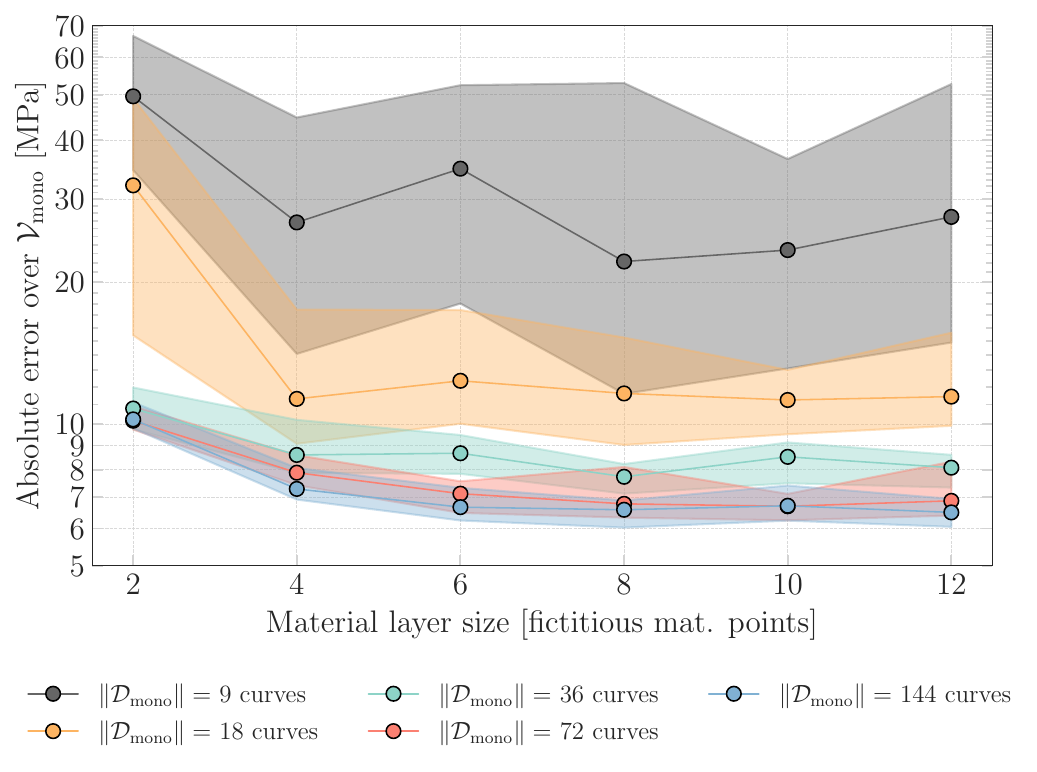}}
\hfill
\subfloat[$\mathcal{V}_{\textrm{prop. GP}} =\{$100 proportional GP-based loading paths$\}$]{\label{fig:valgp}\includegraphics[height = 5.8cm]{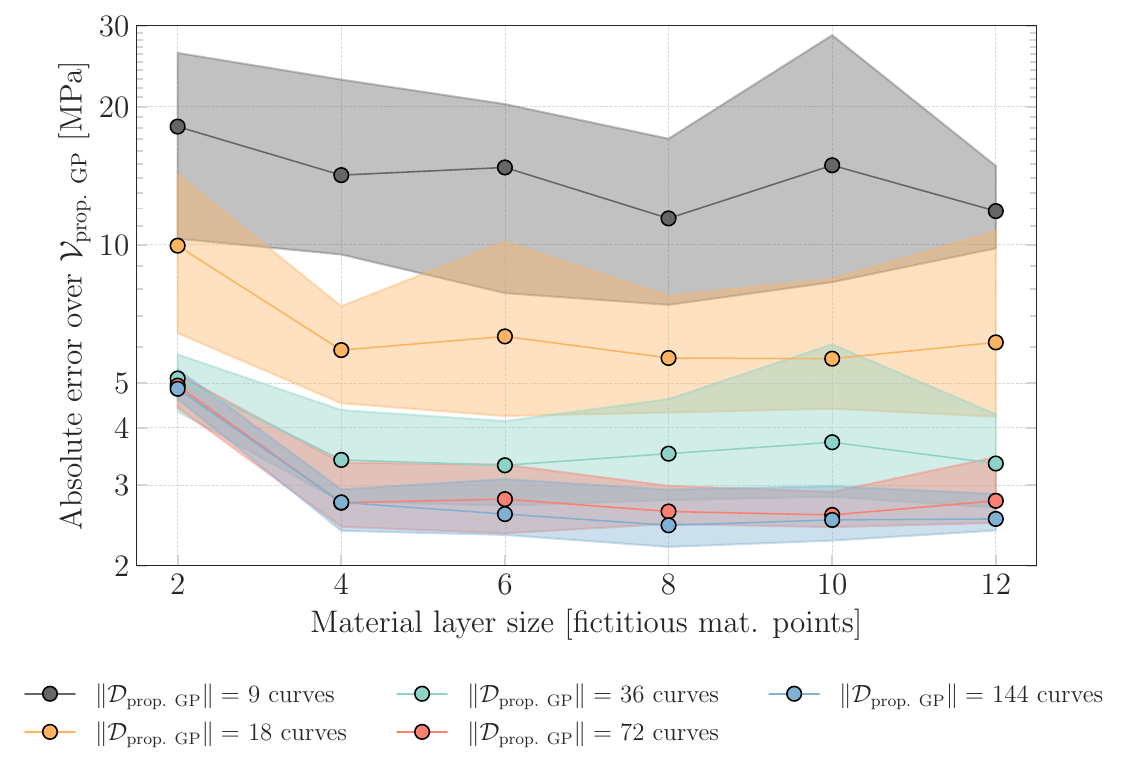}}
\hfill \strut
\caption{Envelope of highest and lowest errors in logarithmic scale from 10 initializations of PRNNs trained on different types of loading and material layer sizes over validation set $\mathcal{V}_{(\cdot)}$. Solid lines with markers correspond to the average validation errors.}
\label{fig:validationdiftraining}
\end{figure}

The second model selection step is focused on finding an optimal size for the material layer. For this purpose, the material layer size is varied considering a range of different training set sizes. Note that at this stage there is no direct comparison between the two training strategies since their training and validation are of matching types. Similarly to the plot in \Cref{fig:decoders}, in  \Cref{fig:validationdiftraining} we show the envelope of best and worst performances, along with the average absolute errors over the validation set $\mathcal{V}_{(\cdot)}$, which this time consists of either monotonic or proportional GP-based curves. In the following sections, we use the PRNNs with material layer size of 8 for both cases, which corresponds to the point where errors are either the lowest among all training sets or have negligible difference with respect to larger layer sizes. 

\subsection{Monotonic loading}
\label{subsec:testmono}
\begin{figure}[h!]
\hfill
\begin{minipage}[b]{.67\textwidth}
\subfloat[Envelope of highest and lowest errors in log scale from 10 initializations of PRNNs trained on different types of loading over $\testmono$\label{fig:enverrormono}]{\includegraphics[width=\textwidth]{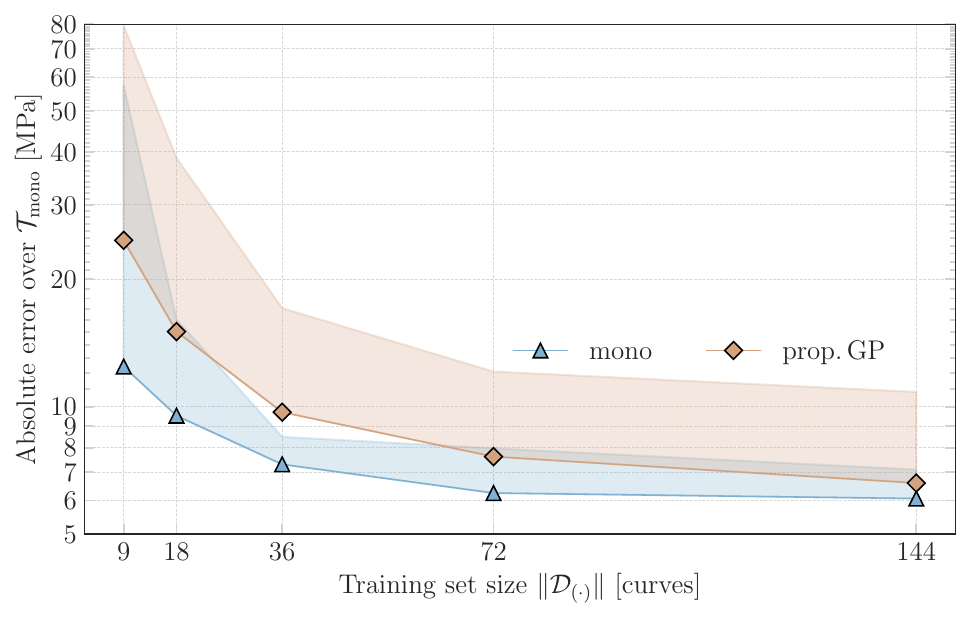}}
\end{minipage}
\hfill
\begin{minipage}[b]{.29\textwidth}
\subfloat[Errors per component for best PRNNs trained on 72 curves over $\testmono$\label{fig:errormonopercomp}]{\includegraphics[width=\textwidth]{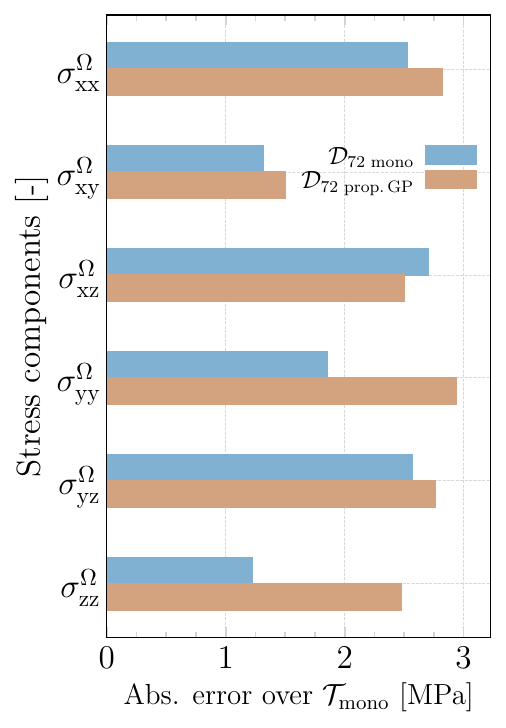}}
\end{minipage}
\hfill \strut
\caption{Envelope of absolute errors from 10 PRNNs trained on different sets and evaluated on test set $\testmono = \{$100 monotonic curves$\}$ on the left and absolute errors per component using the best performing networks with 72 curves on the right. Solid lines with markers correspond to the best performances of each training loading type for several training set sizes.}
\label{fig:testmono}
\end{figure}

As first test scenario, we consider a test set $\testmono$ consisting of 100 monotonic curves in random and unseen directions (type I). We evaluate the networks trained on monotonic (type I) and GP-based paths (type IV) over that test set for different training set sizes. Fig. \ref{fig:enverrormono} shows the lowest absolute and relative errors for both strategies, along with the envelope of absolute errors from 10 initializations. As more data is considered, the error bounds shrink and an optimal training set size can be identified around 72 curves. Although the difference in the lowest errors is still significant, \qty{6.2}{M\pascal} (\qty{5.4}{\%}) vs \qty{7.6}{M\pascal} (\qty{6.3}{\%}), more data translates into marginal gain to both. In the breakdown of the error per component in \Cref{fig:errormonopercomp}, the largest differences in the accuracy are in the $\sigma_\text{yy}^{\Omega}$ and $\sigma_\text{zz}^{\Omega}$ components. The overall performance gap between the two training strategies in this scenario is expected since we are testing on the same loading behavior used to generate the training data of one of the strategies. Another aspect to be considered is the fact that the proportional GP-based curves reach lower strain ranges compared to the monotonic paths for the same number of time steps and step size.  

\begin{figure}[h!]
\centering
\includegraphics[width=.7\textwidth]{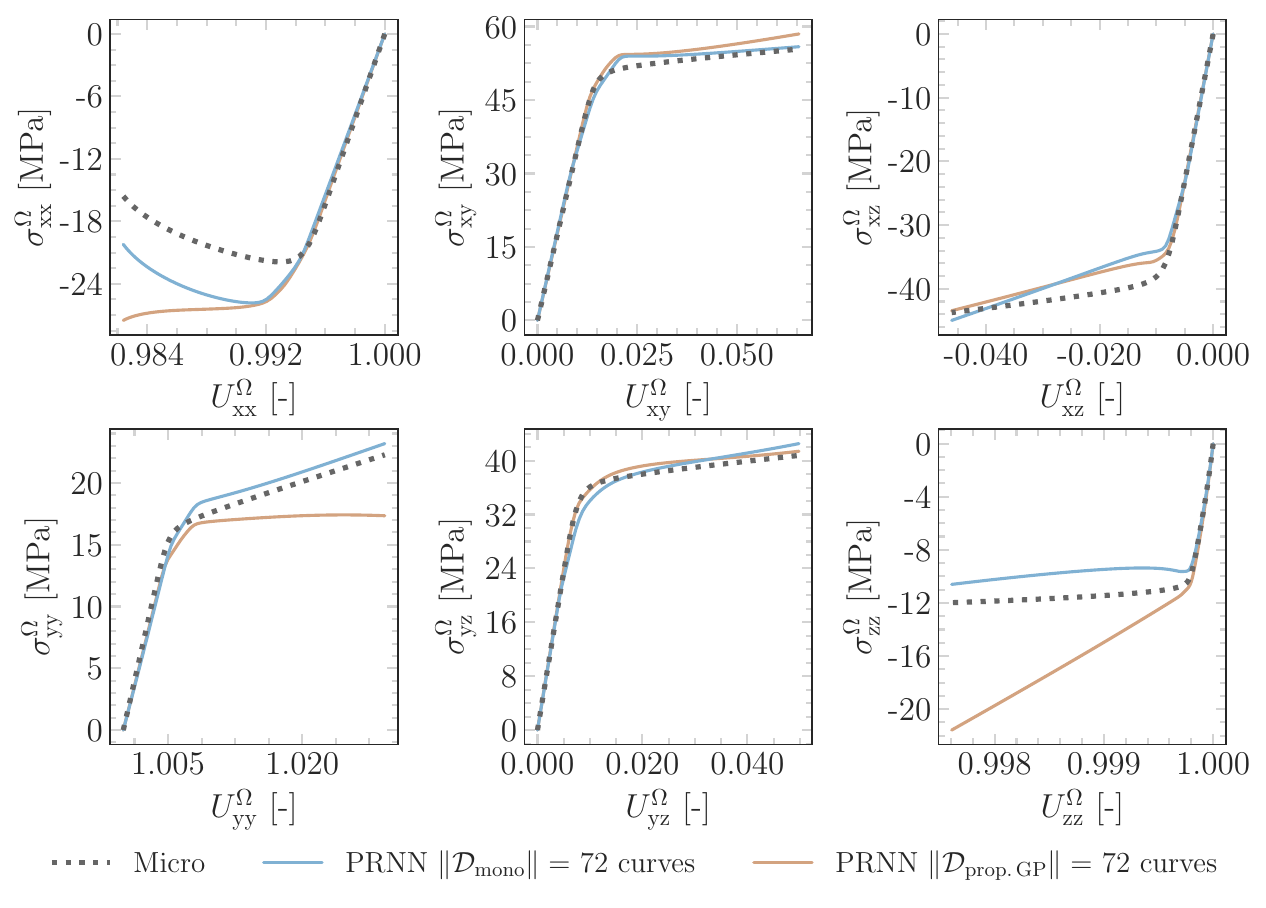}
\caption{Best PRNNs trained on monotonic and GP-based curves on representative curve from test set $\testmono$.}
\label{fig:predmono}
\end{figure}

To illustrate the difference in performance, we select a curve from $\testmono$ with an absolute error close to the best performances from both training strategies. In this case, the prediction error on the curve shown in \Cref{fig:predmono} is around \qty{5.5}{M\pascal} and \qty{7.2}{M\pascal} for the networks trained on monotonic and proportional GP-based curves, respectively. Note that the accuracy loss stands out more in the components with lower stress magnitude such as $\sigma_\text{xx}^{\Omega}$ and $\sigma_\text{zz}^{\Omega}$. An explanation for that comes from the choice of the loss function, the mean squared error. Recall that although normalization of the outputs is considered to balance the difference between stress magnitudes among the components, the MSE remains an absolute metric error. As such, values on the higher end of the normalized range can still dominate the loss, leading to a better fit. Neverthess, satisfatory agreement is observed in the remaining components with the network trained on monotonic data, while the network trained on proportional GP-based curves shows more significant errors. 

\subsection{Monotonic loading with different strain-rates}
\label{subsec:difsr}
Next, we test the ability of the PRNN to capture rate-dependency. For that, two new test sets are considered, $\testslower$ and $\testfaster$, with 100 curves each again in unseen directions (type II). In the first one, the time increment $\Delta t$ is set to be \qty{100}{} times larger than the reference one (\qty{1}{\s}) used for generating the monotonic curves for training, and in the second, the time increment is \qty{100}{} times smaller. The best performances from the 10 PRNNs trained on different types and numbers of curves are summarized in \Cref{tab:errorsdifsr}. Again, the slight advantage of the networks trained on monotonic curves is expected since the loading function in both test sets remains monotonic and reaches similar strain levels. As a result, networks trained with proportional GP-based curves show greater benefit from larger sets, as was the case in the previous assessment. Similarly, since the gain is still relatively small compared to doubling the training set size, we continue the analysis with the smaller set for both types.  

\begin{figure}[!h]
\centering
\includegraphics[width=.7\textwidth]{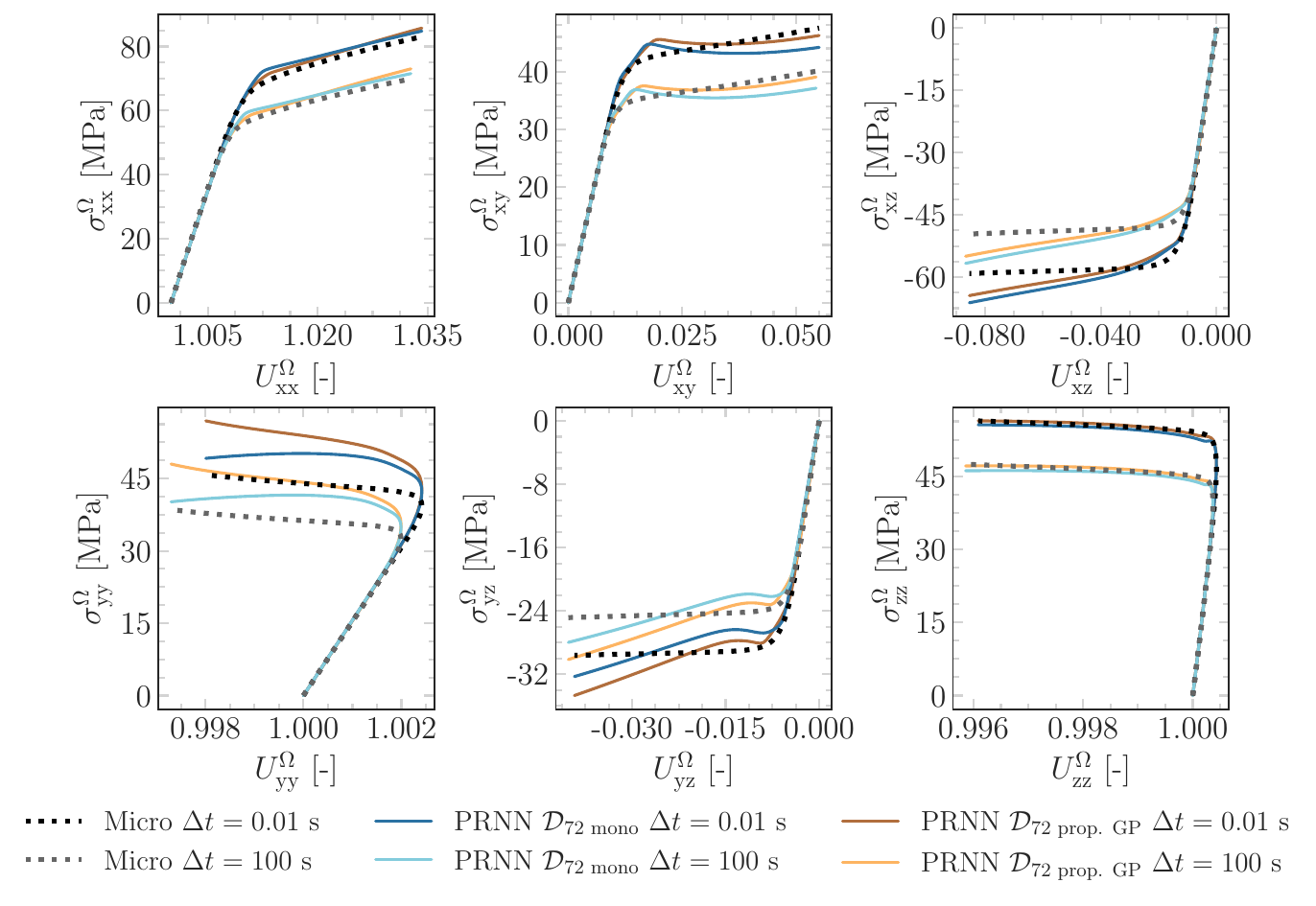}
\caption{Performance of PRNNs trained on monotonic and proportional GP-based curves on test sets with strain-rates 100 times slower and 100 times faster than the one used to create the monotonic training data.}
\label{fig:preddifsr}
\end{figure}

To illustrate the rate-dependent behavior, we use the best networks over each of the test sets and select a representative curve from them to visualize the effect of the different strain-rates (see \Cref{fig:preddifsr}). This is an important milestone of this contribution, especially considering that these strain rates are far from the reference values considered to generate the monotonic curves. The rate dependency in this case is a natural outcome of the elastoviscoplastic model used in the material layer. Encoding rate-dependence in the material layer allows for reproducing this effect without training for it. This is most evident from the error values reported in \Cref{tab:errorsdifsr}, where the test errors are of similar magnitude for test sets with unseen strain-rates as for the test set with the same strain rate as used for the training data. In contrast to modern RNNs, our latent variables have physical interpretation, and, more importantly, evolve according to the same physics-based assumptions considered in the micromodel.  
\begin{table}
\centering
\caption{Summary of lowest absolute errors from 10 PRNNs trained on different types of curves over test sets $\testmono$, $\testfaster$ and $\testslower$}
\begin{tabular}{rcccc}
\toprule
Training loading type & \multicolumn{2}{c}{Monotonic} & \multicolumn{2}{c}{Prop. GP} \\ 
Training set size & 72 & 144 & 72 & 144 \\
\midrule
Abs. error over $\testmono$ [$\unit{M\pascal}$] & 6.2 & 6.1 & 7.6 & 6.6 \\
Abs. error over $\testfaster$ [$\unit{M\pascal}$] & 6.6 & 6.5 & 7.5 & 6.7 \\
Abs. error over $\testslower$ [$\unit{M\pascal}$] & 5.7 & 5.5 & 7.1 & 6.2 \\
\bottomrule
\end{tabular}
\label{tab:errorsdifsr}
\end{table}
 
\subsection{Unloading/reloading behaviour}
\label{subsec:unloading}

In this section, three types of unloading/reloading paths are tested with data sets from type III, IV and V. In all cases, every scenario is assessed based on a test set with 100 curves. Networks trained with both training strategies (based on type I and type IV curves) are evaluated.

\subsubsection{Predefined unloading/reloading function}
\label{subsubsec:propunl}

\Cref{tab:errorunlfixed} presents the lowest error from 10 networks over the test set of proportional curves with pre-defined unloading $\testunlfixed$ (type III). It can be observed that both training strategies lead to similar performances. Note that although the networks trained on proportional GP-based curves can still benefit from a larger training set, we continue the experiments with 72 curves as the gain in accuracy from doubling the training set is minimal. It is also interesting how the networks trained on monotonic paths are still slightly more accurate than the ones that have been trained with unloading. We see this as a result of two subtle advantages: (i) a loading/unloading test function much similar to the monotonic loading paths, especially the first half of the curves in $\testunlfixed$, than to the arbitrary unloading in the proportional GP-based curves and (ii) the time increment in the test curves are the same as the ones in the monotonic curves. 

\begin{figure}[!h]
\centering
\includegraphics[width=.7\textwidth]{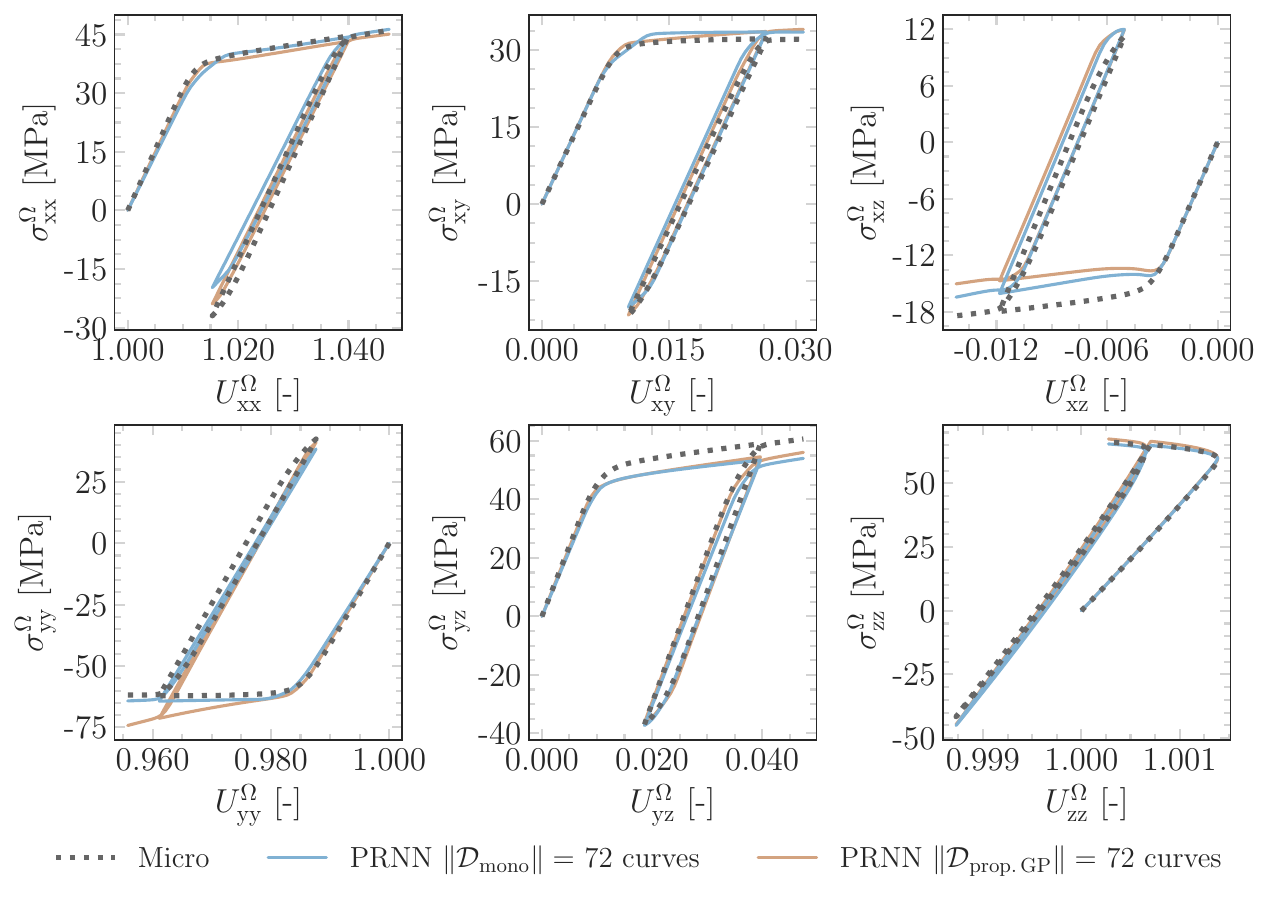}
\caption{Best PRNNs trained on monotonic and proportional GP-based curves on representative curve from test set $\testunlfixed$.}
\label{fig:predunl}
\end{figure}

While these aspects help elucidate the on par performance, they do not express their significance. These networks have never seen any sort of unloading in training  but are still quite capable of extrapolating to such behavior, correctly accounting for the effect of the plastic deformation. This corroborates the findings in \cite{Maiaetal2023}, where a path-dependent material model in the material layer allowed path-dependency to arise naturally. Here, we verify that the method is general and can be extended to account for other non-linearities and time dependencies. \Cref{fig:predunl} shows the predictions on a curve from $\testunlfixed$ with representative errors using the best performing network. Note how close the predictions are to each other and the good agreement with respect to the micromodel solution. 
\begin{table}[!h]
\centering
\caption{Summary of lowest absolute errors from 10 PRNNs trained on different types of curves over test set $\testunlfixed$}
\begin{tabular}{rcccc}
\toprule
Training loading type & \multicolumn{2}{c}{Monotonic} & \multicolumn{2}{c}{Prop. GP} \\ 
Training set size & 72 & 144 & 72 & 144  \\
\midrule
Abs. error over $\testunlfixed$ [MPa] & 6.7 & 6.8  & 7.0 & 6.5  \\
\bottomrule
\end{tabular}
\label{tab:errorunlfixed}
\end{table}

\subsubsection{Proportional and random unloading/reloading}
\label{subsub:propgpbased}

\begin{figure}[!h]
\centering
\subfloat[Proportional GP-based curve from test set $\testunlgpbased$\label{fig:predunlgpbased}]{\includegraphics[width=.65\textwidth]{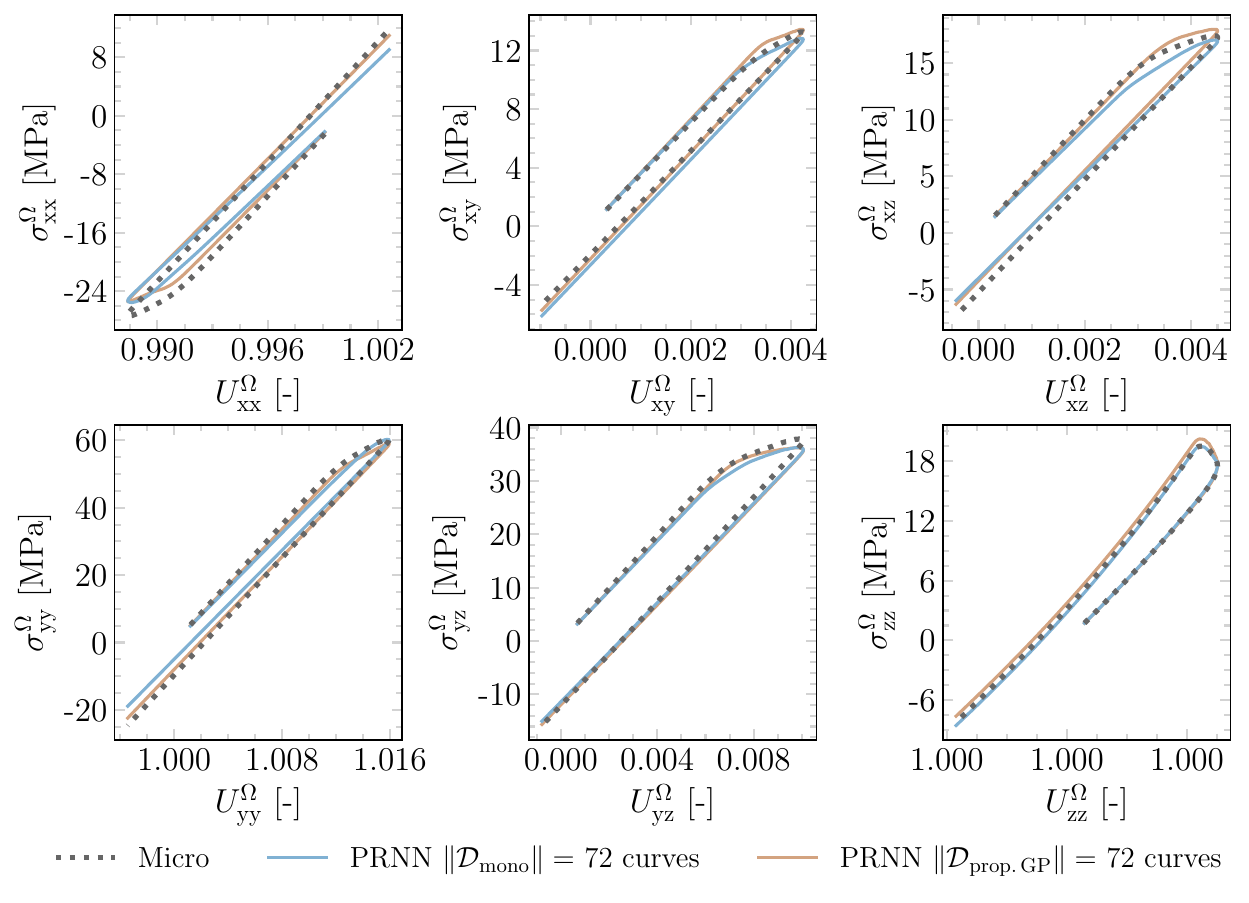}}
\subfloat[Error per component over $\testunlgpbased$\label{fig:errorunlgpbasedpercomp}]{\includegraphics[width=.315\textwidth]{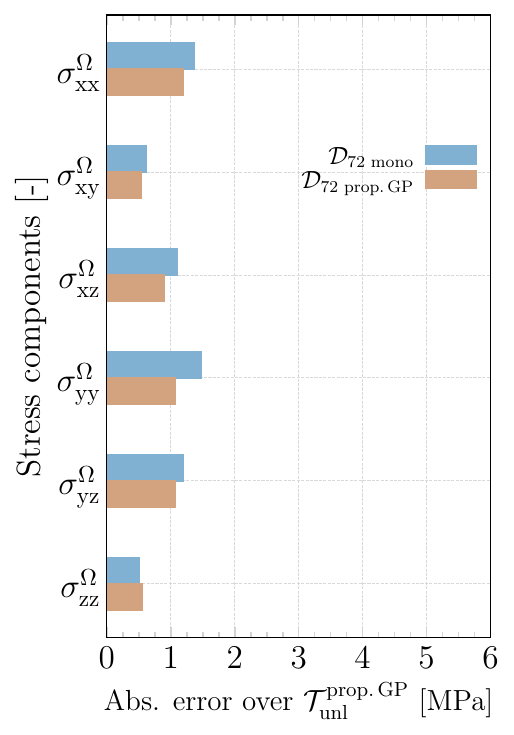}}
\hfill
\subfloat[Non-proportional GP-based curve from test set $\testunlgp$\label{fig:predunlgp}]{\includegraphics[width=.65\textwidth]{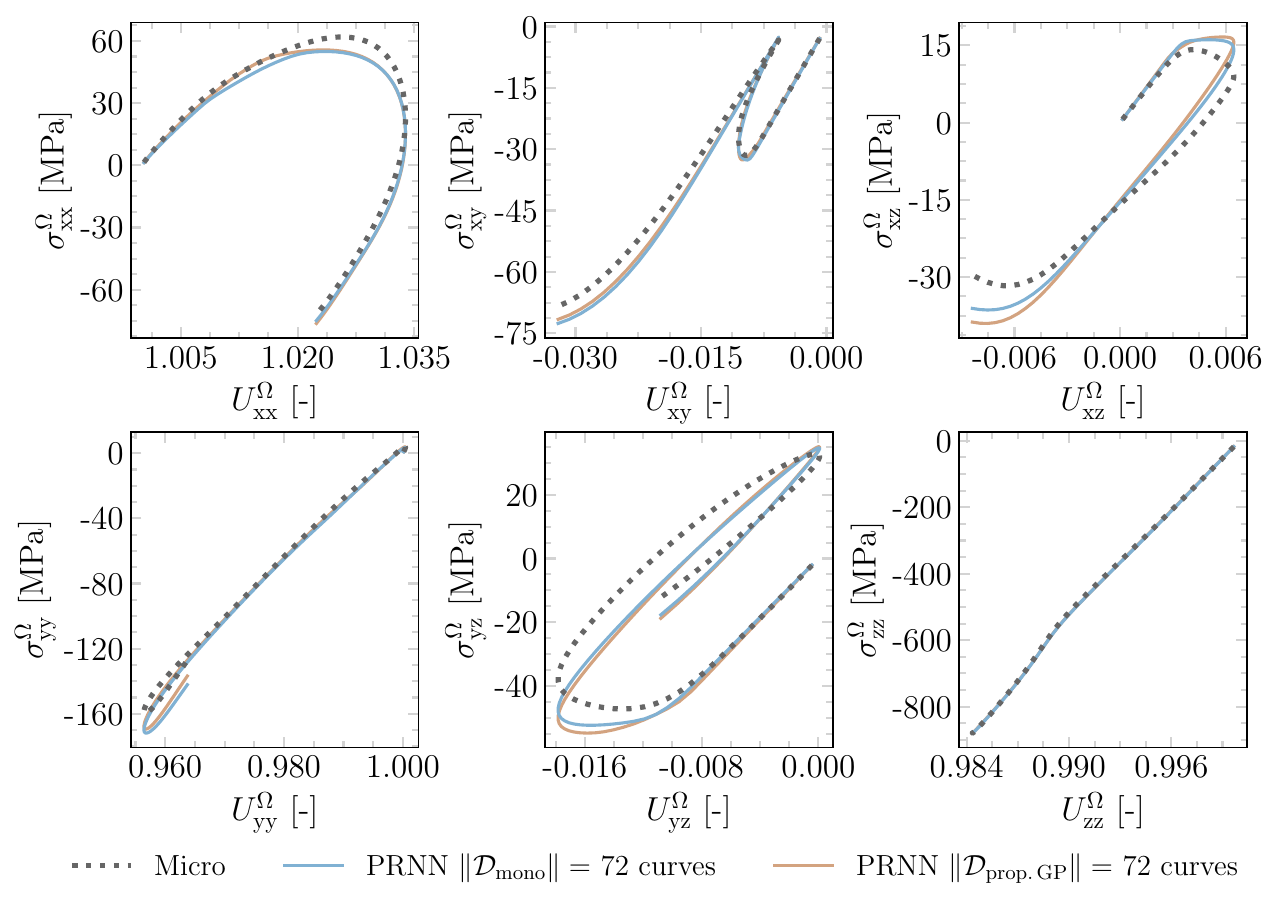}}
\subfloat[Error per component over $\testunlgp$\label{fig:errorunlgppercomp}]{\includegraphics[width=.315\textwidth]{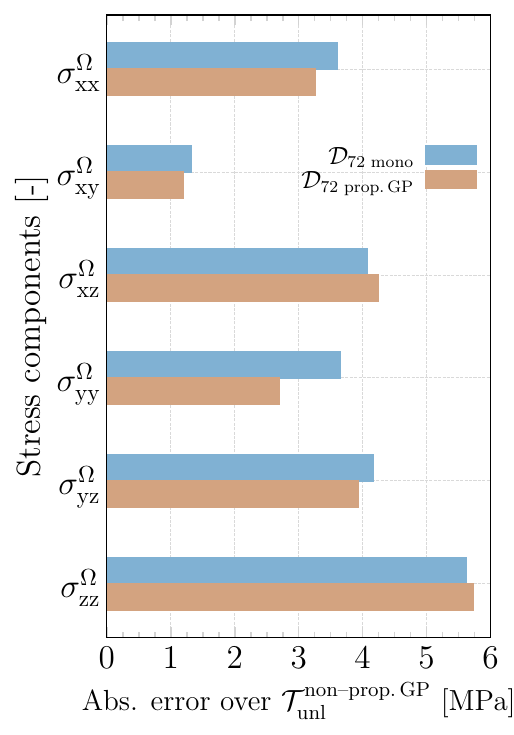}}
\caption{Best PRNNs trained on monotonic and GP-based curves on representative curves from two different test sets with random unloading/reloading.}
\label{fig:predrandomunl}
\end{figure}

In this experiment, the test set $\testunlgpbased$ is used to represent loading paths with unloading-reloading taking place at random times. These curves consist of the same type of loading used in one of the training strategies, which is similar to the situation discussed in Section \ref{subsec:testmono}. Naturally, this results in lower test errors compared to the networks trained on monotonic loading paths, as shown in \Cref{tab:errorgps}. To illustrate the best performance among the 10 networks considered for each strategy, we select a curve $\testunlgpbased$ with errors close to the average lowest absolute error (see \Cref{fig:predunlgpbased}), along with the errors per component (see \Cref{fig:errorunlgpbasedpercomp}).
\begin{table}
\caption{Summary of lowest absolute and relative errors from 10 PRNNs trained on different types of curves over test sets $\testunlgpbased$ and $\testunlgp$}
\centering
\begin{tabular}{rcccc}
\toprule
Training loading type & \multicolumn{2}{c}{Monotonic} & \multicolumn{2}{c}{Prop. GP} \\ 
Training set size & 72 & 144 & 72 & 144  \\
\midrule
Abs. (and rel.) error over $\testunlgpbased$ [MPa] (\%) & 3.2 (8.1) & 3.4 (7.8) & 2.7 (5.6) & 2.6 (5.1)  \\
Abs. (and rel.) error over $\testunlgp$ [MPa] (\%) & 11.5 (3.4) & 12.2 (3.6) & 11.0 (3.1) & 10.9 (3.0) \\
\bottomrule
\end{tabular}
\label{tab:errorgps}
\end{table}

\subsubsection{Non-proportional and random unloading/reloading}
\label{subsub:nonpropgpbased}
For the last part of the experiments on the accuracy of the network, the test set $\testunlgp$ is considered. Curves from this set have more complex unloading behavior and significantly higher stress levels compared to the proportional paths in $\testunlgpbased$. This time, the slight gain in accuracy shown in \Cref{tab:errorgps} from training with the proportional non-monotonic data is examined along with the relative errors. This way, we verify that although the absolute test errors have increased, the performances remain consistent with the values seen so far (below \qty{10}{\%}) in terms of relative errors. 

In \Cref{fig:predunlgp}, a representative curve from $\testunlgp$ illustrates the best performance of both strategies over this set. The difficulty in predicting the lowest magnitude stress (in this case, $\widehat{\sigma}_\text{xz}^{\Omega}$) becomes more evident, as well as the variety of unloading, which this time is different in each of the components. While some components go through unloading (\textit{e.g.} $\widehat{\sigma}_\text{xx}^{\Omega}$ and $\widehat{\sigma}_\text{xy}^{\Omega}$), others are monotonically increasing (\textit{e.g.} $\widehat{\sigma}_\text{zz}^{\Omega}$) and reaching high stress levels, which are naturally followed by higher absolute errors per component as seen in \Cref{fig:errorunlgppercomp}.

\begin{figure}[!h]
\centering
\subfloat[Stress-time and stress-strain views of component $\sigma_\text{yz}^{\Omega}$\label{fig:predunlgpbasedstresstime}]{\includegraphics[width=.5\textwidth]{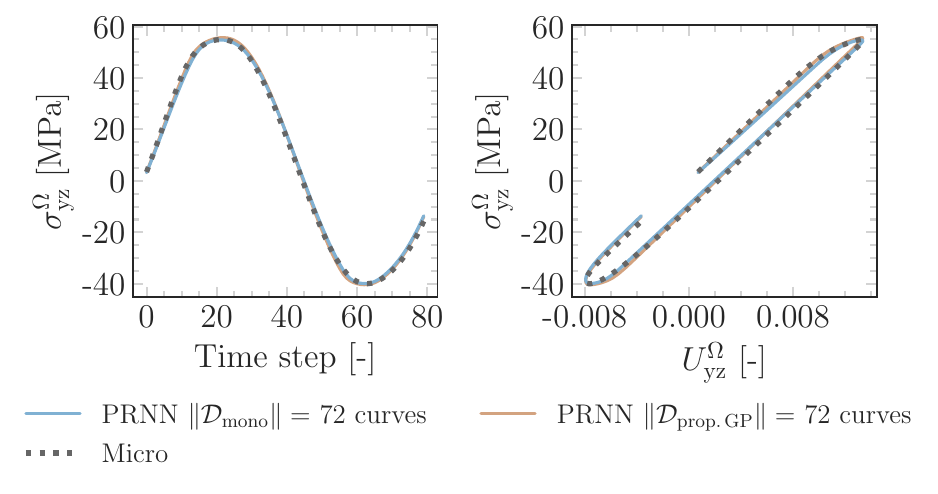}}
\hfill
\subfloat[Stress-time and stress-strain views of component $\sigma_\text{zz}^{\Omega}$\label{fig:predunlgpstresstime}]{\includegraphics[width=.5\textwidth]{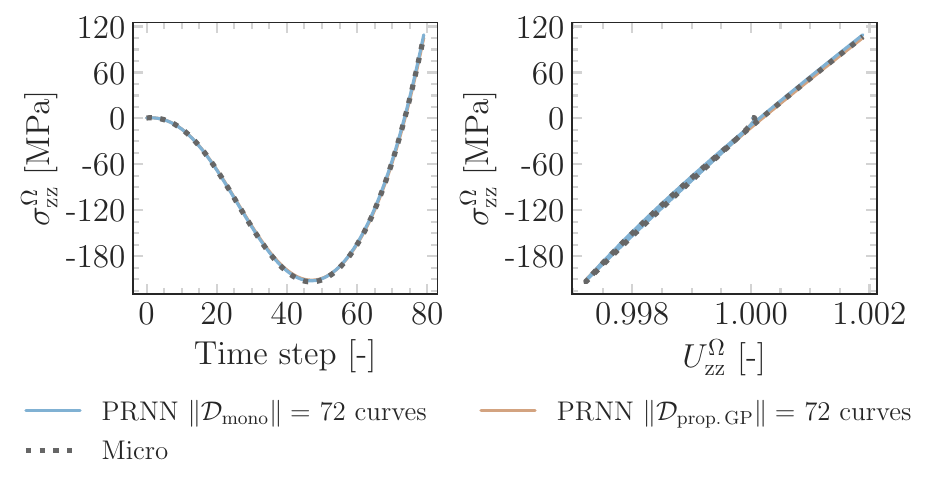}}
\caption{Orthotropic behavior of selected components in loading path from $\testunlgp$.}
\label{fig:predrandomunlstresstime}
\end{figure}

An additional curve from $\testunlgp$ is selected and shown in \Cref{fig:predrandomunlstresstime} to highlight another aspect not yet discussed, the orthotropic behavior of the micromodel. Note that the unloading in the $z$-direction follows the same stress-strain path as the loading, indicating that the elastic fiber is acting as the main load-bearing component. In contrast, the shear stress in $yz$ follows unloading in a different branch due to the development of plastic strains in the matrix. 

Finally, although training with monotonic curves showed consistent and relatively accurate responses in most scenarios, the random and smooth type of loading paths explored in the previous and the current section are deemed to be more general and better representative of arbitrary functions. In both cases, training with 72 proportional GP-based curves has shown better performance and is therefore used to assess the network's capabilities in \Cref{sec:applications}, where the network is used as a material model in several applications. 

\section{Runtime comparison}
\label{sec:speedupgp}
In this section, we perform a runtime comparison to assess the speed-up of the proposed approach in terms of the homogenized stress evaluation. For that purpose, we continue with the loading type investigated in \Cref{subsub:nonpropgpbased} (type V), and select one model from the 10 initializations trained on 72 proportional GP-based curves to represent the best overall performance. Here, we use the network with the lowest error over $\testunlgpbased$. The choice could also have been based on $\testunlgp$, but in favor of simplicity, in a case where the experiments presented in \Cref{sec:numexp} are not carried out, choosing from $\testunlgpbased$ implies a simpler model selection based on a single type of loading. 
\begin{figure}[!h]
\centering
\includegraphics[width=.12\textwidth]{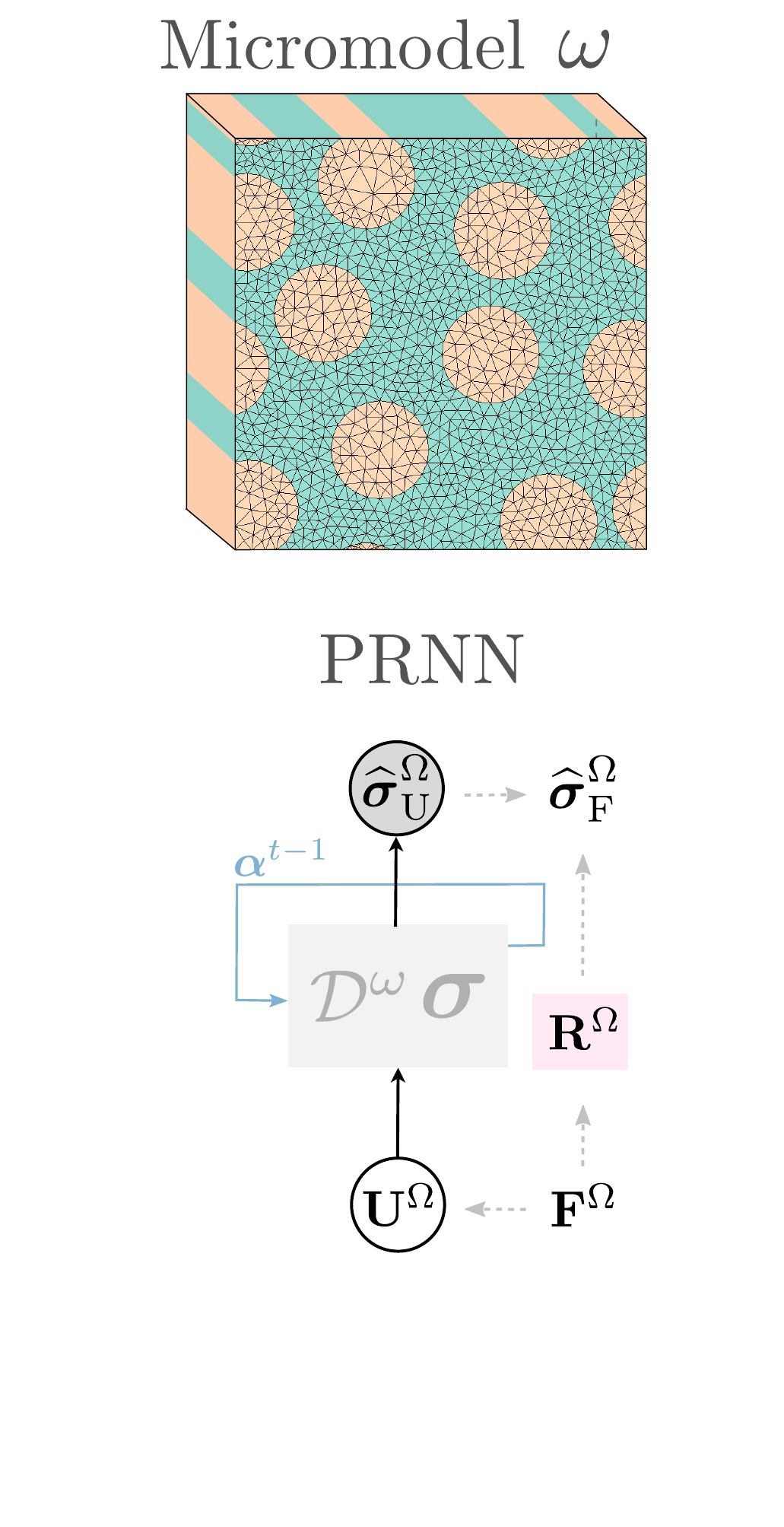}
\includegraphics[width=.7\textwidth]{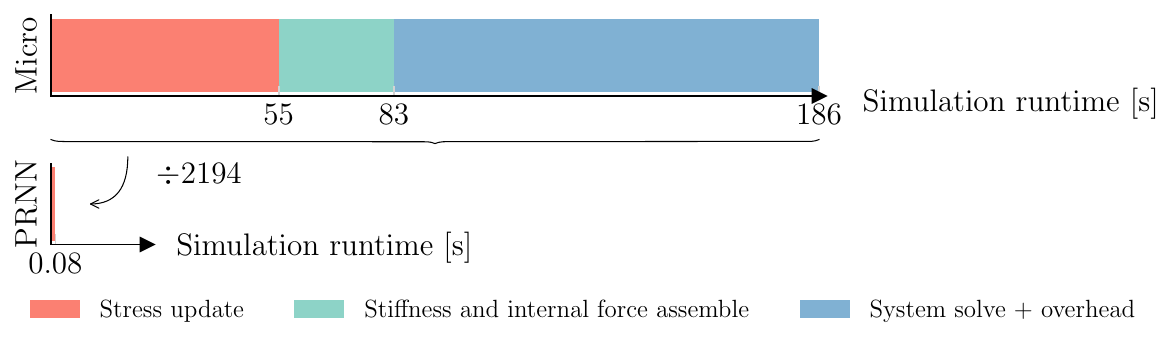}
\caption{Breakdown of simulation runtime using the micromodel and the PRNN averaged over 150 type V loading paths.}
\label{fig:speedup}
\end{figure}

In this work, all simulations, including the data generation and training procedure for the network, were executed on a single core of a Xeon E5-2630V4 processor on a cluster node with 128 GB RAM running CentOS 7. Because we are interested in the final homogenized stress $\boldsymbol{\sigma}_\text{F}^{\Omega}$, we include in the PRNN runtime, the time spent in the transformations to bring the predicted homogenized stress back to the original frame, as illustrated in \Cref{fig:microprnn}. For this comparison, we use as input the converged strain path and time increments from the micromodel simulations. The micromodel mesh is shown in \cref{fig:rvemesh} and consists of wedge elements integrated with 2 points in the thickness direction, totaling \num{4992} integration points and \qty{7860} degrees of freedom.
 
Averaging over the results from 150 simulations, we break down the runtime from the full-order model in the three main parts depicted in \Cref{fig:speedup}. With the micromodel, roughly \qty{30}{\%} of the simulation is spent evaluating the constitutive models at the integration points, around $\qty{15}{\%}$ goes to the assembly of the global stiffness matrix and internal force vector and more than half of the total time is spent solving the system, totaling 186 seconds. In contrast, the network needs only \qty{0.08}{s} to compute the homogenized stress state, which results in a speed-up of three orders of magnitude when compared to the full-order solution. 

It is worth mentioning that this is only an estimate of the actual speed-up. In the general case, the speed-up depends on several other aspects, such as the robustness of the tangent stiffness matrix, the complexity of the loading case, and the size of the micromodel. In multiscale settings, the gain can be higher since the cost associated with an iteration at the macroscale builds on a much higher execution time when using the micromodel compared to the network. To illustrate that, we include an additional speed-up comparison in the last application of \Cref{sec:applications}. 

\section{Applications}
\label{sec:applications}

In this section, the PRNN trained to surrogate the constitutive behavior of the micromodel in \Cref{sec:numexp} is tested in applications in which its robustness also plays a role in obtaining the equilibrium path. By robustness, we understand the ability of the network to provide not only accurate stress predictions, as verified in Section \ref{sec:numexp}, but also a tangent stiffness matrix that is stable enough for tracing an equilibrium path as close as possible to the one obtained by solving the micromodel. Previously, the entire strain path and time increments obtained from converged micromodel simulations were used as input. Here, the network is directly employed as the material model and therefore the stress prediction at each time step affects the following stress/strain state. In this case, lack of smoothness of the surrogate output may lead the iterative procedure to venture outside the training domain, potentially giving rise to divergence from the true solution.

\begin{figure}[!h]
\centering
\includegraphics[width=.35\textwidth]{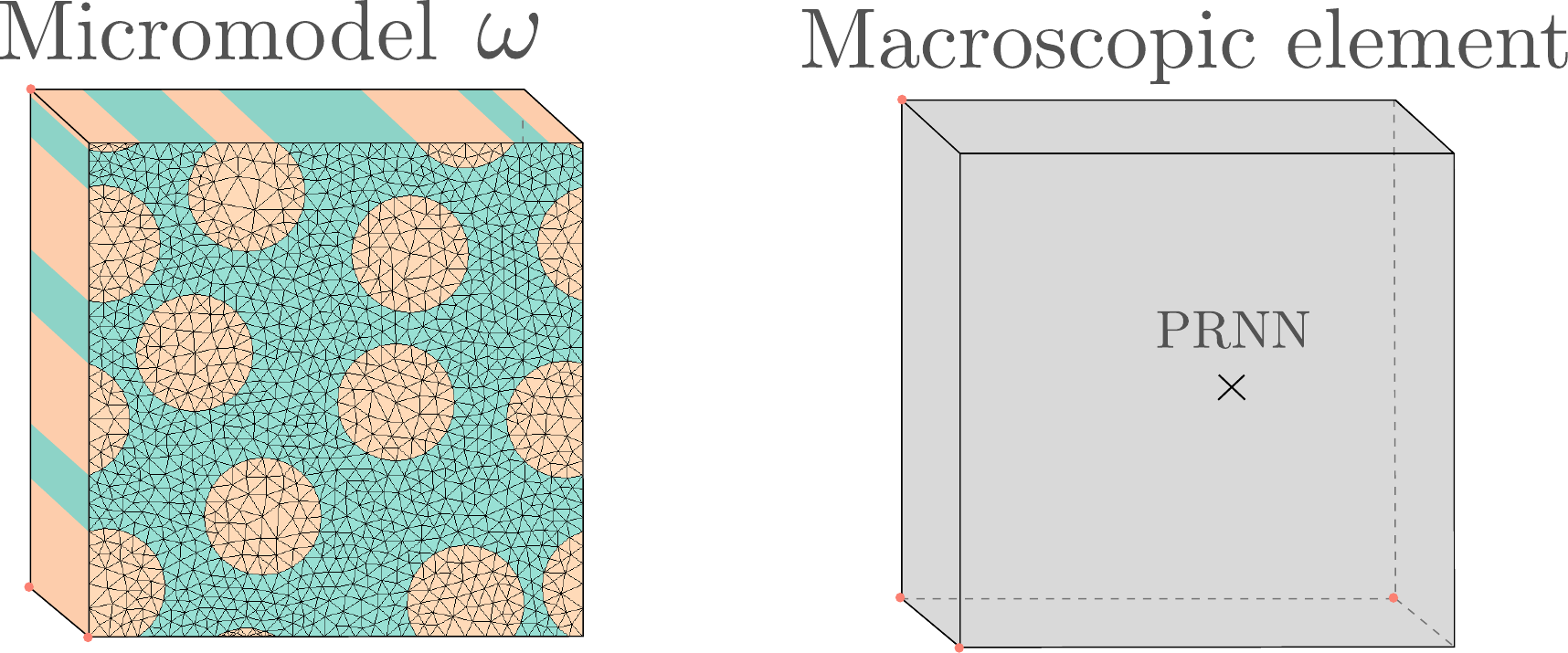}
\caption{Micromodel and PRNN meshes used in the applications.}
\label{fig:meshes}
\end{figure}

For all applications in this section, we use the network with the lowest error on $\testunlgpbased$. This time, to simulate its performance as a surrogate model to the micromodel, the network is embedded in a FE mesh that consists of a single 8 node hexahedral element with the same dimensions as the micromodel and one integration point with constitutive response given by the PRNN, as illustrated in \cref{fig:meshes}. To process the deformation gradient $\mathbf{F}^{\Omega}$ into a simpler input space for the network (\textit{i.e.} $\mathbf{U}^{\Omega}$) and obtain the stresses in their original frame of reference ($\predstressrot$) using $\mathbf{R}^{\Omega}$, we use the scheme in \Cref{fig:microprnn}. For better readability, we drop the subscript, and refer to the final stresses simply as $\widehat{\boldsymbol{\sigma}}^{\Omega}$. Furthermore, for both the micromodel and the hexahedral element, in addition to the constrained displacements to avoid rigid body motion (see \Cref{fig:micropbc}), periodic boundary conditions are applied.

In the first application, we test the ability of the model to reproduce the stress relaxation phenomenon. In the second, we deal with cyclic loading and in the last application, the network is embedded in the general nonlinear framework developed by \citet{Kovacevic2022} to account for off-axis and constant strain-rate loading conditions. For the latter, we also include speed-up measurements to illustrate how aspects such as step size and tangent stiffness smoothness can play a role in increasing or decreasing the speed-up compared to the study in \Cref{sec:speedupgp}. 

\subsection{Relaxation}
\label{subsec:relax}
In this study, a loading function to reproduce the stress relaxation phenomenon is devised. For that, the micromodel and the PRNN are loaded until a given strain level is reached $\boldsymbol{\varepsilon}^{\Omega}_{0}$ at $t = t_0$, when the stress level is $\boldsymbol{\sigma}^{\Omega}_{0}$.  After that, the strain is held constant, while a gradual stress reduction takes place. 
For that, we use the arc-length control introduced in \Cref{sec:datagen} and control the stretching in the $x$-direction, leaving the remaining directions free to deform. 
\begin{figure}[!h]
\hfill
\subfloat[Homogenized stress-time response of micromodel and PRNN on the top and homogenized strain-time on the bottom\label{fig:macrorelax}]{
\includegraphics[width=.5\textwidth]{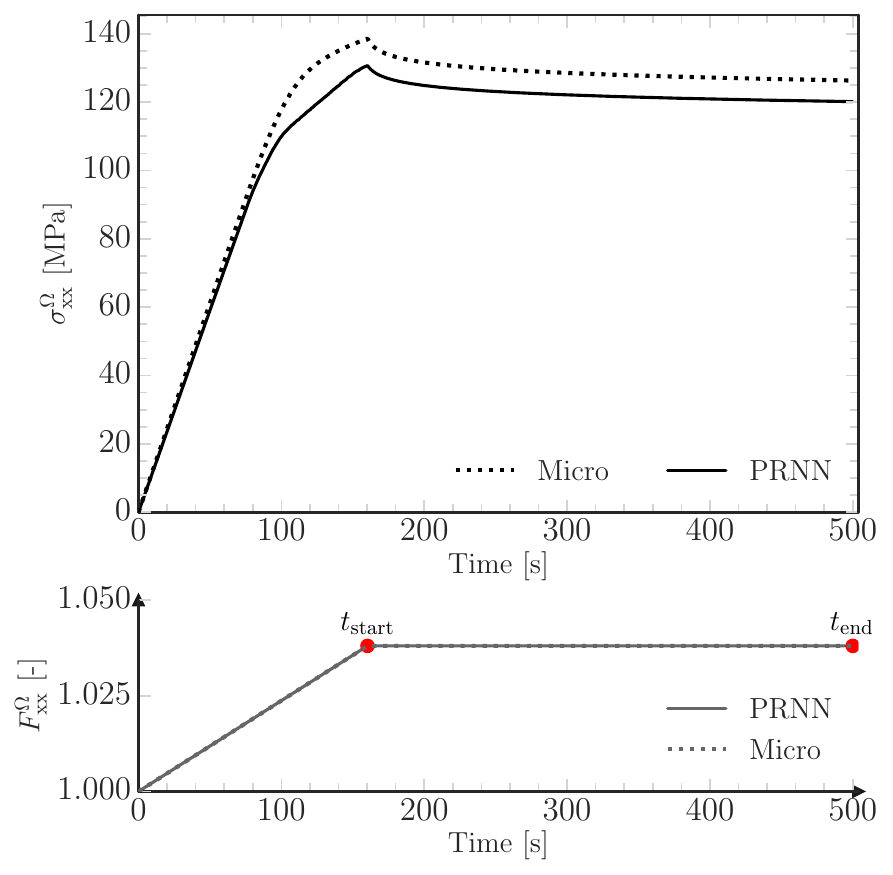}}
\hfill
\subfloat[Micromodel state at the start and end of constant strain loading\label{fig:microrelax}]{\includegraphics[width=.26\textwidth]{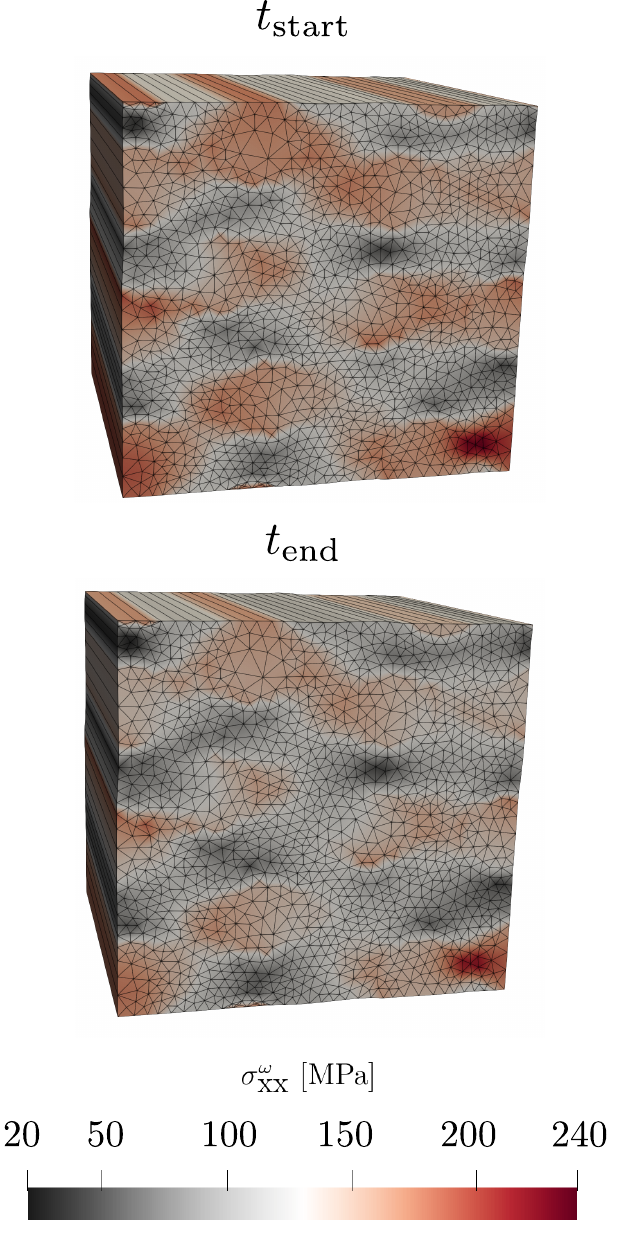}}
\hfill \strut
\caption{Homogenized stress-time response of micromodel and PRNN subjected to uniaxial stretch in $x$  until t = \qty{160}{\s}, when the strain is held constant until the end of the simulation t = \qty{500}{\s}. On the right, the full-field of stresses of the micromodel for the start and end of the constant strain loading (in \emph{red}).}
\label{fig:predrelax}
\end{figure}

In this example, the micromodel and the homogeneous hexahedral element are loaded with $\|\Delta \mathbf{u}^\text{c} \| = \qty{5.e-6}{\mm}$ and $\Delta t = \qty{1}{\s}$ until $t_0 = \qty{160}{\s}$, when the strain level at that point is held constant until the total time of \qty{500}{s} is reached and the analysis is terminated, as depicted in the lower plot of \Cref{fig:macrorelax}. In the upper plot, despite the mismatch in the stress at the start of the constant strain \textrm{plateau}, the overall stress-time response of the micromodel is in relatively good agreement with the network's prediction, with an average error of \qty{6}{MPa} (\qty{5}{\%}). While this case represents a challenging scenario for even modern RNNs due to the long strain repetition, the expected stress decaying behaviour in the prediction comes as an inherent outcome of using a material model that incorporates a spectrum of relaxation times in the material layer. To illustrate the slight difference in the stress state at the beginning and end of the constant strain plateau, we show in \Cref{fig:microrelax} two snapshots of the full-field solution.

\subsection{Cyclic loading}
\label{subsec:cyclic}
\begin{figure}[!h]
\centering
\subfloat[Stress-strain curve with first and last cycle highlighted\label{fig:stressstraincyclic}]{\includegraphics[width=.5\textwidth]{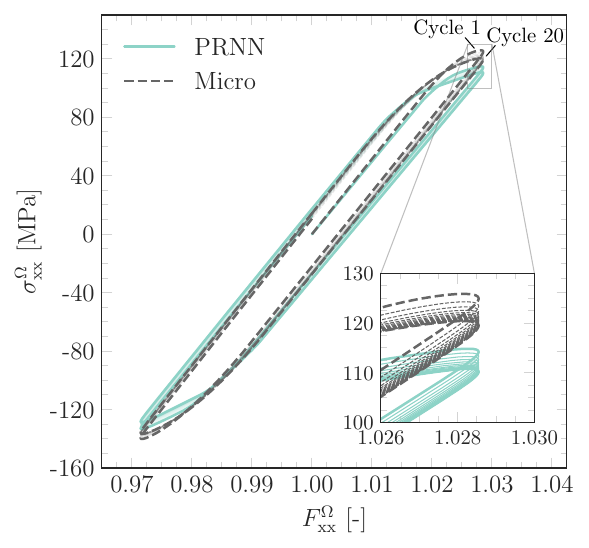}}
\hfill
\subfloat[Peak strains of selected components \label{fig:straintimecyclic}]{\includegraphics[width=.5\textwidth]{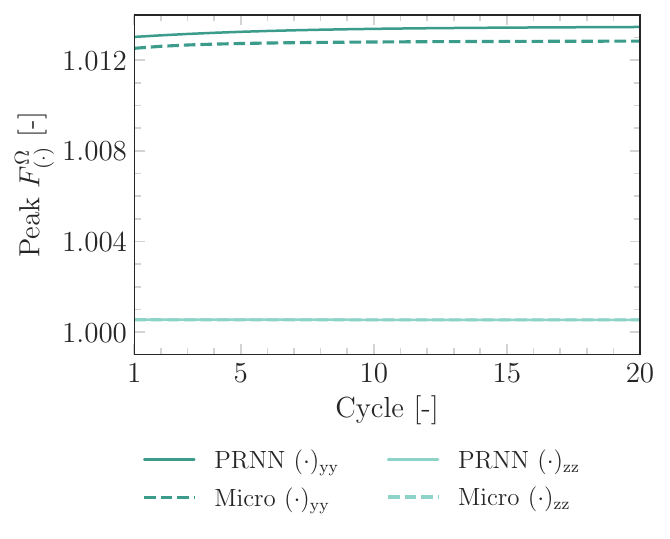}}
\caption{Stress-strain response of micromodel and PRNN subjected to uniaxial cyclic loading.}
\label{fig:predcyclic}
\end{figure}

To assess the network's performance on cyclic loading, we continue with the arc-length method and same boundary conditions as the previous application but now the uniaxial stretch at time $t$ is described as:
\begin{equation}
\label{eq:cyclicloadfunc}
F_\textrm{xx}^{\Omega} = 1 + \frac{\qty{6.e-3}{}}{l} \, \sin{\left(\frac{2\pi}{\num{1000}}\, t\right)}  
\end{equation}
where $t$ is the time step index and $l = \qty{0.021}{mm}$ is the side length of the micromodel. 20 cycles are considered, each consisting of \num{1000} steps with $\Delta t = \qty{1}{\s}$. \Cref{fig:stressstraincyclic} shows the stress-strain curve for the entire loading history. The network reproduces the reverse plasticity and the hysteresis behavior in the cyclic response. Because \Cref{eq:cyclicloadfunc} consists of a symmetric loading with constant peak and valley strains, a slow stress decay over the cycles takes place. This asymptotic relaxation process can be observed in the inset in \Cref{fig:stressstraincyclic} and is of similar nature to the one discussed in Section \ref{subsec:relax}. Overall, good agreement is found between the PRNN and the micromodel solution. This is further assessed by unrolling the stress-strain response in time and extracting the peak and valley quantities. 

\begin{figure}[!h]
\includegraphics[trim = 0 0 0 165, clip, width=\textwidth]{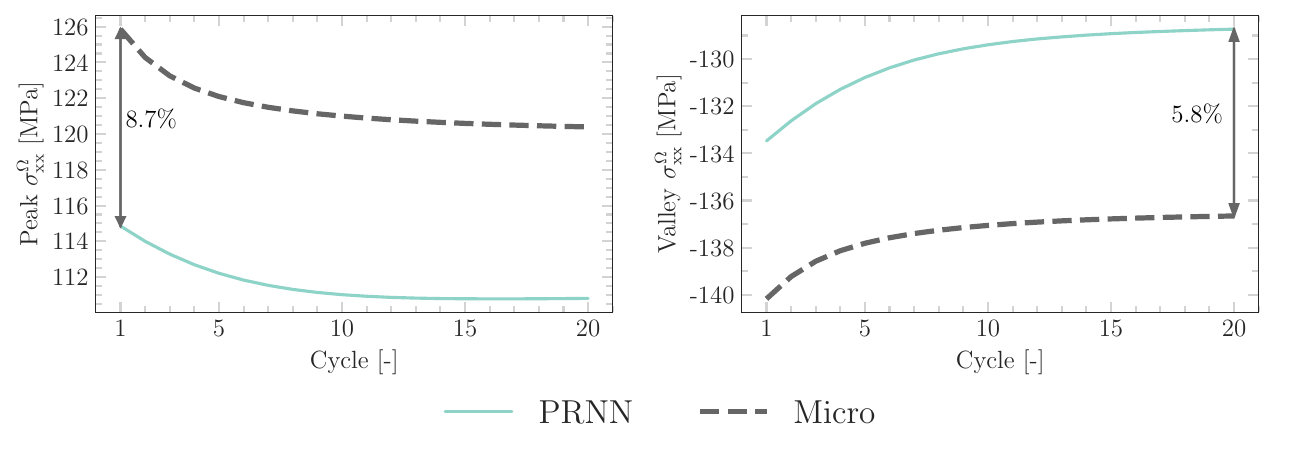}
\caption{Evolution of maximum and minimum stresses for all cycles with double arrows marking the relative error corresponding to the highest absolute difference between the micromodel and PRNN subjected to uniaxial cyclic loading.}
\label{fig:predcyclicstresstime}
\end{figure}

First, the peak strain values from the diagonal components not controlled by the arc-length are plotted in \Cref{fig:straintimecyclic}. In this case, the strain path obtained by the network remains close to the true solution and only minor deviations are observed in the $F_\textrm{yy}$ component. Naturally, different loading conditions lead to different levels of accuracy of the strain paths due to the indirect displacement control equation considered in this work. As for the stresses, the envelopes of maximum and minimum values for the entire loading history are shown in \Cref{fig:predcyclicstresstime}. In each, the highest absolute error is marked by double arrows, along with the corresponding relative error. Both absolute and relative errors are within the range of errors obtained in previous sections. 

\subsection{Constant strain-rate under off-axis loading}

\begin{figure}[!h]
\vfill
\subfloat[Global and local coordinate systems]{\includegraphics[width=0.33\textwidth]{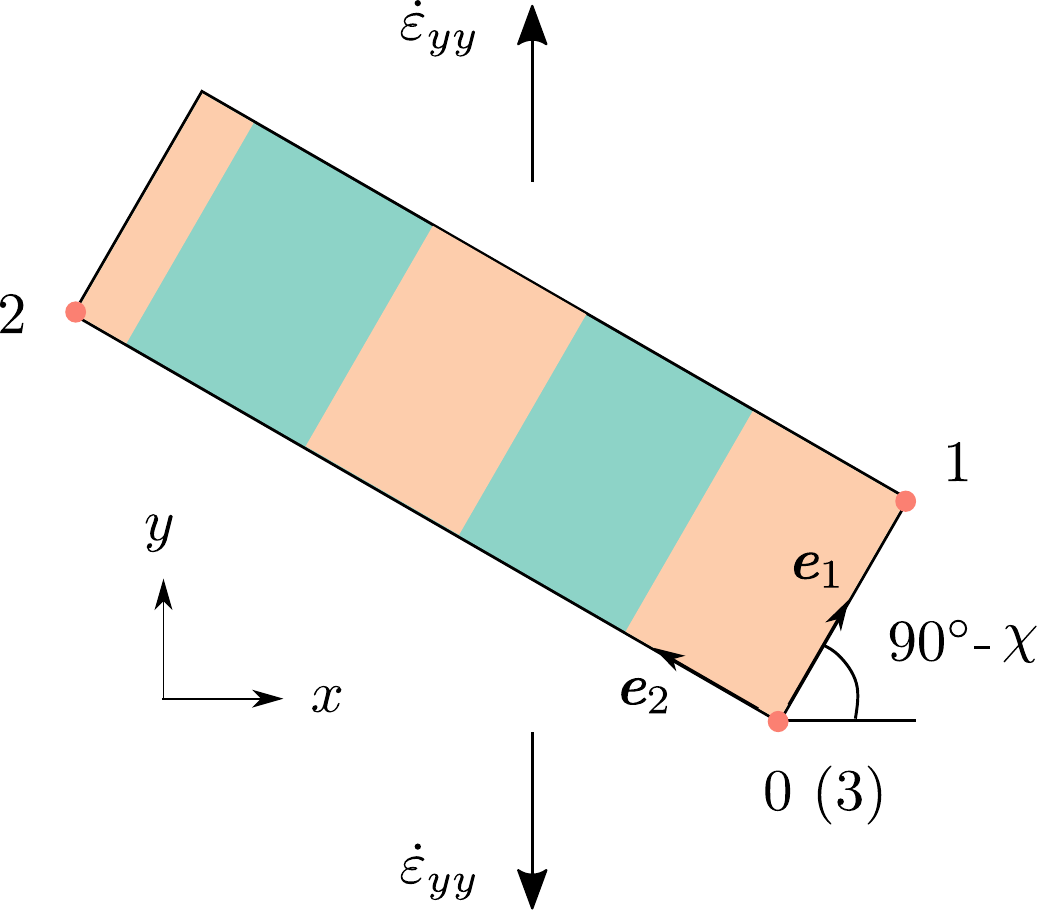}}
\subfloat[Micromodel in the local system]{\includegraphics[width=0.33\textwidth]{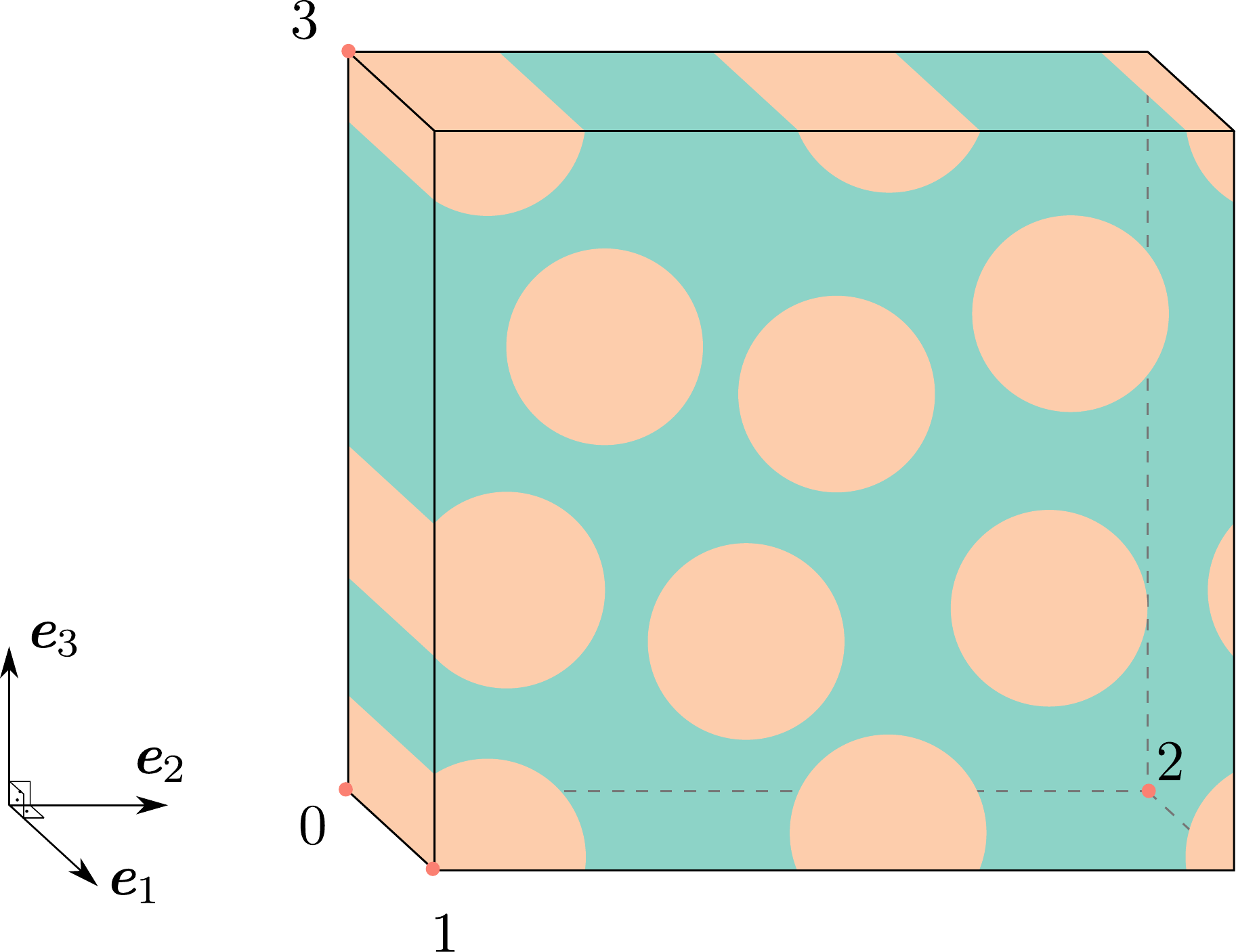}}
\subfloat[Reorientation of micromodel due to applied loading $\dot{\varepsilon}_\text{yy}$\label{fig:csrstates}]{\includegraphics[width=0.33\textwidth]{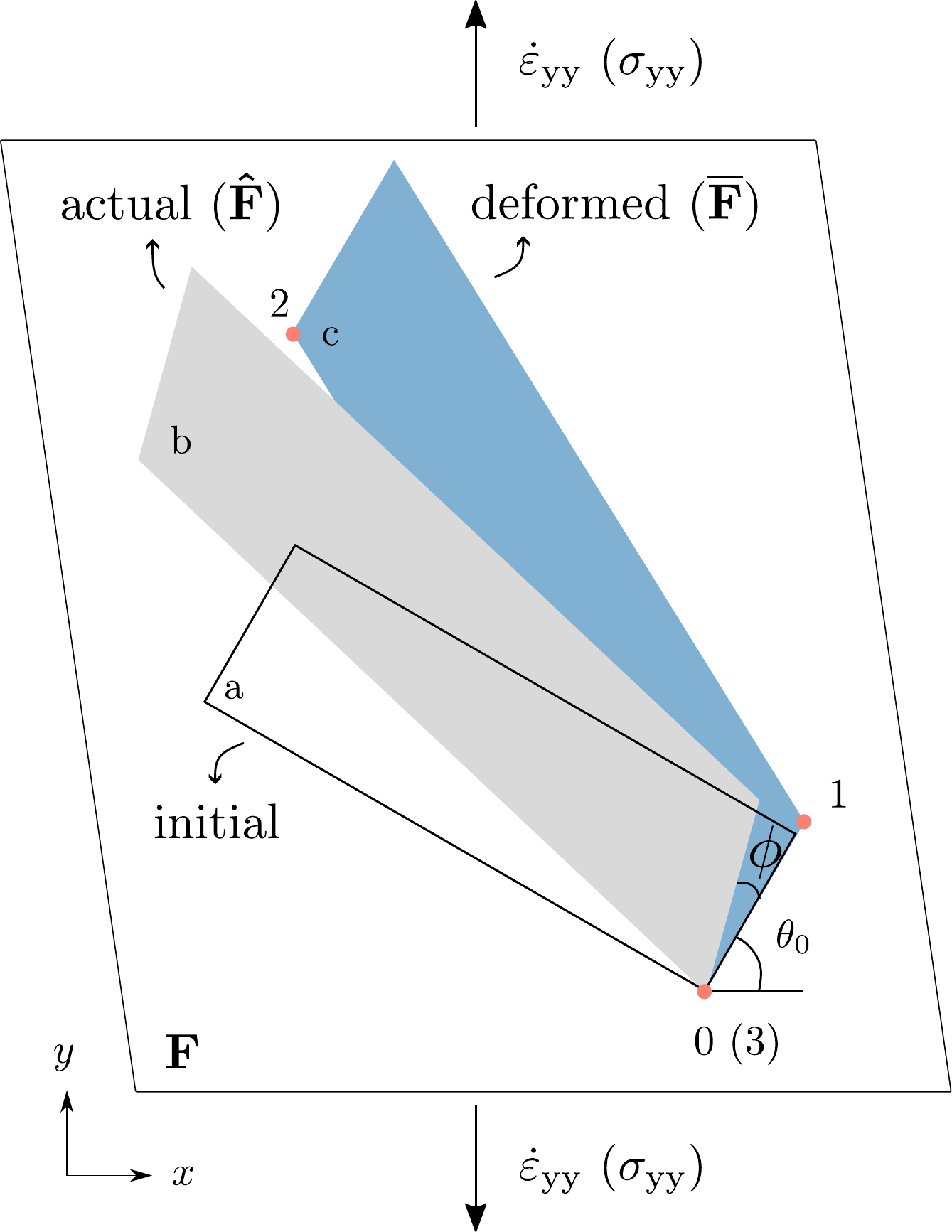}}
\caption{Global and local coordinate systems with imposed strain-rate $\dot{\varepsilon}_\text{yy}$ on $y$ direction and off-axis angle $\chi$ and reorientation of micromodel due to applied loading $\dot{\varepsilon}_\text{yy}$ from initial angle $\theta_0$ to $\theta_1 = \theta_0 + \phi$ based on the deformed state \cite{Kovacevic2022}.}
\label{fig:constsr}
\end{figure}

For the last application, a dedicated strain-rate based arc-length formulation is used to reproduce the response of unidirectional composites subjected to off-axis loading \cite{Kovacevic2022}. In this formulation, two coordinate systems are needed: the global ($x$ and $y$ axes) and the local (1, 2 and 3 axes), as depicted in Fig. \ref{fig:constsr}. In the global coordinate system, the initial fiber orientation with respect to the $y$-axis is defined according to a given off-axis angle $\chi$. The micromodel is then subjected to constant strain-rate ($\dot{\varepsilon}_\textrm{yy}$) under uniaxial stress conditions. With that, equivalent homogenized deformation and stress states need to be derived in the local frame, and the transformations between global and local coordinate systems are taken care by the custom arc-length model. 

In this work, we embed the network in the \emph{local} frame, with the time increment $\Delta t$ and the homogenized deformation gradient $\overline{\mathbf{F}}$ as input and the homogenized stress $\overline{\boldsymbol{\sigma}}$ as the output. In \Cref{fig:csrstates}, we show the three relevant configurations in this framework. In the simulation, due to the applied loading, the micromodel edge $0\text{--}1$ tied to the local axis $\boldsymbol{e}_1$ should rotate with an angle $\phi$ with respect to the initial configuration (from ``a" to ``b"), going from the initial angle $\theta_0$ to a new angle $\theta_1 = \theta_0+\phi$. However, to avoid rigid-body rotation of the RVE, the controlling node 1 is fixed in the shearing direction, but the angle $\phi$ is implicitly taken into account through the constraint equation and the unit force vector of the arc-length model. For that reason, configuration ``c", in which $\boldsymbol{e}_1$ is always aligned to the initial fiber orientation, is used to evaluate $\phi$. 

In the local frame, the homogenized deformation gradient $\mathbf{\overline{F}}$ is given by:
\begin{equation}
\label{eq:homstrainlocalcsr}
\overline{\mathbf{F}} =
\begin{bmatrix}
\overline{F}_{11} & \overline{F}_{12} & 0 \\
0 & \overline{F}_{22} & 0 \\
0      & 0      & \overline{F}_{33} \\
\end{bmatrix}
\end{equation}
To ensure the global constant strain rate condition, a special constraint equation $g$ derived by equating the homogenized deformation gradient component in the global frame $F_\text{yy}$ to the value imposed from the input is considered:
\begin{equation}
\label{eq:arclengthcsr}
g = \underbrace{\overline{F}_{11} \sin (\theta_0) \sin (\theta_1) + \overline{F}_{22} \cos (\theta_0) \cos (\theta_1) + \overline{F}_{12} \cos (\theta_0) \sin (\theta_1)}_{F_\text{yy}\text{ calculated from micromodel}} - \underbrace{\exp (\varepsilon_\text{yy}^{t-1} + \dot{\varepsilon}_\text{yy} \Delta t}_{F_\text{yy}\text{ imposed from input}}) = 0
\end{equation}
where $\varepsilon_\text{yy}^{t-1}$ is the total strain in the global loading direction from the last converged time step. Another vital part of the framework is related to the update on the unit force vector applied at the controlling nodes. In this case, the geometrically nonlinear effect on the unit force vector comes not only from the change in configuration ``a" to ``c" but also from the change in orientation of the micromodel that $\phi$ introduces. Finally, to relate the stresses from both frames, one can use the load factor $\lambda$ from the arc-length formulation, which is equivalent to the $\sigma_\text{yy}$ stress component in the global frame, to transform it to the local frame:
\begin{equation}
\label{eq:homstresslocalcsr}
\overline{\boldsymbol{\sigma}} = \sigma_\text{yy} \ \begin{bmatrix}
\sin^2 (\theta_1) & \cos (\theta_1) \ \sin (\theta_1) & 0 \\
\cos (\theta_1) \ \sin (\theta_1) & \cos^2 (\theta_1) & 0 \\
0      & 0      & 0 \\
\end{bmatrix}
= 
\begin{bmatrix}
\sigma_{11} & \sigma_{12} & 0 \\
\sigma_{21} & \sigma_{22} & 0 \\
0      & 0      & 0 \\
\end{bmatrix}
\end{equation}

\begin{figure}[!h]
\centering
\includegraphics[width=0.65\textwidth]{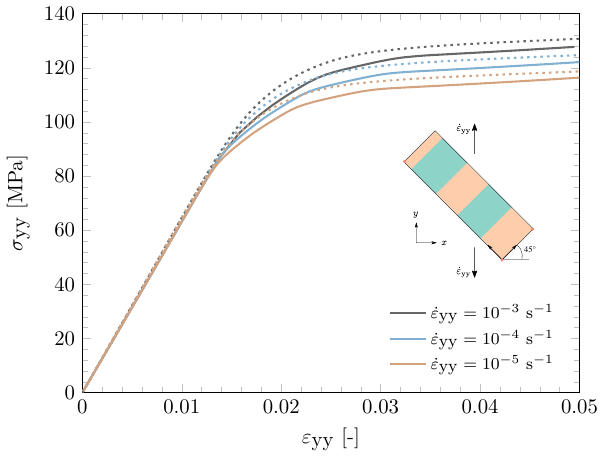}
\caption{Global stress-strain curve from off-axis composite with $\chi = 45^{\circ}$ and different strain-rates $\dot{\varepsilon}_\text{yy}$. Solid and dashed lines refer to the micromodel solution and the PRNN prediction, respectively.}
\label{fig:globalstressconstsr}
\end{figure}

\begin{figure}[!ht]
\centering
\subfloat[Diagonal components]{\label{fig:locstressdiag} \includegraphics[width=0.5\textwidth]{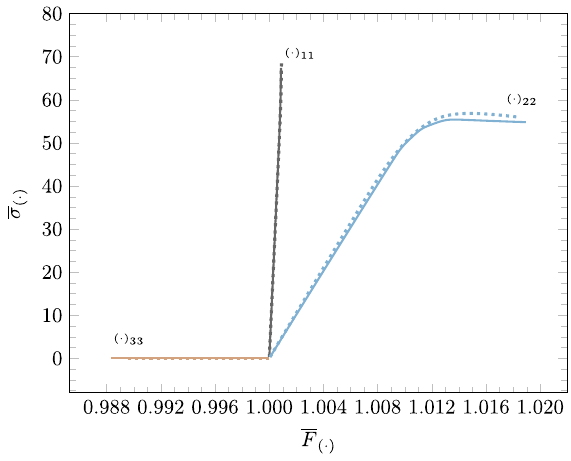}}
\subfloat[Off-diagonal component]{\label{fig:locstressoffdiag}\includegraphics[width=0.5\textwidth]{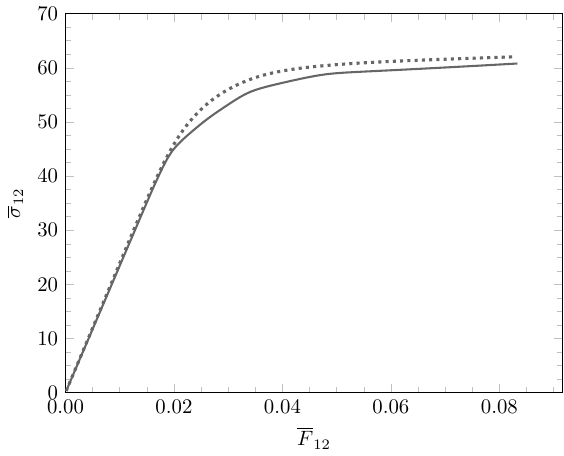}}
\caption{Stress and deformation in the local system for $\chi = 45^{\circ}$ and $\dot{\varepsilon}_\textrm{yy} = 10^{-4} \, \unit{\per\s}$. Solid and dashed lines refer to the micromodel solution and the PRNN prediction, respectively.}
\label{fig:localstressCSR}
\end{figure}

In this contribution, we particularise the framework to $\chi = \qty{45}{\degree}$ and strain-rates $\dot{\varepsilon}_\text{yy} = [\num{e-5} \, \unit{\per\second}, \num{e-4} \, \unit{\per\second}, \num{e-3} \, \unit{\per\second}]$, resulting in three simulations in total. For more details on the formulation and derivation of the expressions presented in this section, the reader is referred to \cite{Kovacevic2022}. Starting with the global stress-strain response, results in \Cref{fig:globalstressconstsr} show satisfactory agreement with the full-order solution. This is yet another verification of the capability of the network to handle rate-dependency. We also inspect in \Cref{fig:localstressCSR} the evolution of separate pairs of stress and deformation gradient components in the local frame. It is emphasized, that in this simulation, none of these stress and strain components is directly controlled since there is a nonlinear relation where the evolution of the load in local frame depends on the computed deformation, except for the $\bar\sigma_{33}$ which is kept at zero. It is observed that all deformation and stress components computed with the PRNN remain close to those coming from the micromodel.

A final assessment is made in terms of speed-up. This time, because an adaptive stepping scheme is used, the termination criterion (maximum norm) can be reached with a different number of macroscopic steps depending on the tangent stiffness matrix. For that reason, in \Cref{tab:speedupdifsr}, in addition to the breakdown of the total simulation time into the three tasks shown in \Cref{fig:speedup} and the speed-up, we also show the number of steps. With less iterations, speed-ups range from \num{2900} to \num{4000}, which is significantly higher than the one obtained in \Cref{sec:speedupgp} ($\approx\num{2200}$), where neither the adaptive stepping scheme nor the network's tangent and predictions are used to define the next step in tracing the equilibrium path. Other aspects, such as macroscopic mesh density and algorithmic parameters, can also influence the speed-up and the relative times of each task with respect to the total time. In this particular case with a single macroscopic element, using the PRNN as the homogenized constitutive model means that most of the time is dedicated to evaluating the network. With that, we demonstrate the potential of the proposed approach as a robust and efficient model in a practical application. 
\begin{table}
\caption{Breakdown of simulation time and speed-up for different strain-rates $\dot{\varepsilon}_\text{yy}$ and $\chi = 45^{\circ}$, each averaged over 10 simulations}
\centering
\begin{tabular}{rcccccc}
\toprule
$\dot{\varepsilon}_\text{yy} \, [\unit{\per\second}]$ & \multicolumn{2}{c}{$\num{e-5}$}  & \multicolumn{2}{c}{$\num{e-4}$} & \multicolumn{2}{c}{$\num{e-3}$} \\
Type of analysis & Micro & PRNN & Micro & PRNN & Micro & PRNN \\
\midrule
$N_\text{steps}$ [-] & 293 & 95 & 290 & 95 & 279 & 65 \\
Stress evaluation [s] (\%) & 165 (15) & .222 (59) & 160 (15) & .219 (59) & 160 (14) & .168 (60)  \\
Stiff. and int. force assemble [s] (\%) & 91.6 (8) & .0117 (3) & 89.4 (8) & .0116 (3) & 91.4 (8) & .00875 (3) \\
System solve + overhead [s] (\%) & 843 (77) & .142 (38) & 843 (77) & .142 (38) & 872 (78) & .105 (37) \\
Total simulation time [s] & 1099 & .375 & 1092 & .373 & 1123 & .282 \\
Speed-up [-] & \multicolumn{2}{c}{2929} & \multicolumn{2}{c}{2932} & \multicolumn{2}{c}{3980}  \\
\bottomrule
\end{tabular}
\label{tab:speedupdifsr}
\end{table}

\section{Concluding remarks}
\label{sec:conclusion}

A novel Physically Recurrent Neural Network (PRNN) architecture has been developed to accelerate the microscale analysis of path and rate-dependent heterogeneous materials. The formulation follows the core idea in \cite{Maiaetal2023}, where the homogenized response of a micromodel is obtained by a network with constitutive models embedded in one of its layers. In this \emph{material layer}, we have \emph{fictitious material points} with the same constitutive models and properties as used in the micromodel. The values passed from encoder to the material layer are interpreted as (fictitious) local strains, which are input to the constitutive model assigned to the material points, yielding (fictitious) local stresses. These local stresses are subsequently transformed by a decoder to obtain the homogenized stress.  

What distinguishes the present methodology from the state-of-the-art surrogate models, particularly the ones based on RNNs, is the strong physics-based assumptions built into the model. Here, history-dependency is a natural outcome of the embedded material models. This is because, in addition to the local stress, the material model assigned to a fictitious material point is also in charge of updating its own internal variables (if any), which are stored from one time step to another. Therefore, PRNNs naturally inherit rich memory mechanisms from the constitutive models, bypassing the need to learn these latent dynamics from data.

While the concept of having few fictitious material points representing the homogenized response of a micromodel remains at the core of the method, a new architecture is required to extend the applicability of the network to 3D problems in a finite strain framework. Among the key changes compared to \cite{Maiaetal2023} are the use of the polar decomposition theorem and the principle of material objectivity. With the former, the deformation gradient can be uniquely decomposed into two tensors, namely stretch and rotation. The network is then used to learn the mapping between stretch and unrotated stress, from which the stress in the global coordinate frame is retrieved using the principle of material objectivity. 

For the numerical examples, we considered a unidirectional composite micromodel with rate-dependent plasticity in the matrix and hyperelasticity in the fibers. Two different training strategies (monotonic vs non-monotonic) were considered. When creating the monotonic curves, a single value of time increment was considered so that we could clearly illustrate the exceptional ability of the network to extrapolate to strain rates far from the ones seen during training. We have also tested the performance of the network on curves with increasingly complex unloading behavior. In this case, although the networks trained on monotonic data could capture unloading behavior and performed well in most of the considered scenarios, training on non-monotonic curves led to better performance overall. Comparing the number of curves of the network selected for the numerical applications with previous developments \cite{Maiaetal2023}, now we need twice as many curves to train a PRNN that is twice as big. This linear scaling should not be expected given the exponential increase nature from the curse of dimensionality, yet we can still achieve it.

In \Cref{sec:applications}, we shifted our focus to applications where the PRNN is directly replacing the micromodel in the solution of the equilibrium problem. In the first application, we demonstrated that the network can reproduce relaxation, which can be a difficult behavior to capture with RNNs due to the long repetition of the input (\textit{i.e.} constant strain). In our case, since the constitutive models in the network have such behavior in their formulation, the homogenized response also reflected it robustly. In the second example, cyclic loading was considered, again showing the ability of the network to extrapolate to loading conditions and direction different than those trained for. For the last application, the special arc-length formulation proposed in \cite{Kovacevic2022} to account for off-axis loading and constant strain-rate conditions was employed. We particularised the framework to one off-axis angle and three different strain-rates and showed good agreement with the actual micromodel for a case where the network and its tangent are used to compute the solution of a nonlinear problem. 

To assess the network's potential to accelerate micromodel simulations, we investigated two scenarios. Firstly, the network was used to predict stresses based on the converged strain paths from 150 micromodel simulations, leading to a stress evaluation 2200 faster compared to the full-order model. Then, we assessed the speed-up on a problem in which the PRNN was directly involved in tracing the solution. In that case, the constant strain-rate application was used as a reference. It was observed that the lower number of steps needed when using the PRNN as the material model led to speed-ups even higher, between \num{2900} and \num{4000} for the different strain-rates. In summary, the proposed network provides an efficient model that can describe the rate-dependent, orthotropic response of thermoplastic composites in large deformations. Trained on data generated with a micromodel, the PRNN response remains close to that of the micromodel for a wide range of loading scenarios, including those outside the training range.

\section*{Acknowledgements}
The authors acknowledge the TU Delft AI Initiative for their support through the SLIMM AI Lab. FM acknowledges financial support from the Dutch Research Council (NWO) under Vidi grant 16464.

\biboptions{sort&compress}
\bibliographystyle{elsarticle-num-names} 
\bibliography{references.bib}

\begin{thebibliography}{31}
\expandafter\ifx\csname natexlab\endcsname\relax\def\natexlab#1{#1}\fi
\providecommand{\url}[1]{\texttt{#1}}
\providecommand{\href}[2]{#2}
\providecommand{\path}[1]{#1}
\providecommand{\DOIprefix}{doi:}
\providecommand{\ArXivprefix}{arXiv:}
\providecommand{\URLprefix}{URL: }
\providecommand{\Pubmedprefix}{pmid:}
\providecommand{\doi}[1]{\href{http://dx.doi.org/#1}{\path{#1}}}
\providecommand{\Pubmed}[1]{\href{pmid:#1}{\path{#1}}}
\providecommand{\bibinfo}[2]{#2}
\ifx\xfnm\relax \def\xfnm[#1]{\unskip,\space#1}\fi
\bibitem[{Oliver et~al.(2017)Oliver, Caicedo, Huespe, Hernández, and
  Roubin}]{OLIVER2017}
\bibinfo{author}{J.~Oliver}, \bibinfo{author}{M.~Caicedo},
  \bibinfo{author}{A.~Huespe}, \bibinfo{author}{J.~Hernández},
  \bibinfo{author}{E.~Roubin},
\newblock \bibinfo{title}{Reduced order modeling strategies for computational
  multiscale fracture},
\newblock \bibinfo{journal}{Computer Methods in Applied Mechanics and
  Engineering} \bibinfo{volume}{313} (\bibinfo{year}{2017})
  \bibinfo{pages}{560--595}.
  \DOIprefix\doi{https://doi.org/10.1016/j.cma.2016.09.039}.
\bibitem[{Ghavamian et~al.(2017)Ghavamian, Tiso, and Simone}]{GHAVAMIAN2017}
\bibinfo{author}{F.~Ghavamian}, \bibinfo{author}{P.~Tiso},
  \bibinfo{author}{A.~Simone},
\newblock \bibinfo{title}{Pod–deim model order reduction for strain-softening
  viscoplasticity},
\newblock \bibinfo{journal}{Computer Methods in Applied Mechanics and
  Engineering} \bibinfo{volume}{317} (\bibinfo{year}{2017})
  \bibinfo{pages}{458--479}.
  \DOIprefix\doi{https://doi.org/10.1016/j.cma.2016.11.025}.
\bibitem[{Rocha et~al.(2019)Rocha, {van der Meer}, and Sluys}]{ROCHA2019}
\bibinfo{author}{I.~Rocha}, \bibinfo{author}{F.~{van der Meer}},
  \bibinfo{author}{L.~Sluys},
\newblock \bibinfo{title}{Efficient micromechanical analysis of
  fiber-reinforced composites subjected to cyclic loading through time
  homogenization and reduced-order modeling},
\newblock \bibinfo{journal}{Computer Methods in Applied Mechanics and
  Engineering} \bibinfo{volume}{345} (\bibinfo{year}{2019})
  \bibinfo{pages}{644--670}.
  \DOIprefix\doi{https://doi.org/10.1016/j.cma.2018.11.014}.
\bibitem[{Rocha et~al.(2021)Rocha, Kerfriden, and {van der Meer}}]{ROCHA2021}
\bibinfo{author}{I.~Rocha}, \bibinfo{author}{P.~Kerfriden},
  \bibinfo{author}{F.~{van der Meer}},
\newblock \bibinfo{title}{On-the-fly construction of surrogate constitutive
  models for concurrent multiscale mechanical analysis through probabilistic
  machine learning},
\newblock \bibinfo{journal}{Journal of Computational Physics: X}
  \bibinfo{volume}{9} (\bibinfo{year}{2021}) \bibinfo{pages}{100083}.
  \DOIprefix\doi{https://doi.org/10.1016/j.jcpx.2020.100083}.
\bibitem[{Heider et~al.(2020)Heider, Wang, and Sun}]{HEIDER2020}
\bibinfo{author}{Y.~Heider}, \bibinfo{author}{K.~Wang},
  \bibinfo{author}{W.~Sun},
\newblock \bibinfo{title}{So(3)-invariance of informed-graph-based deep neural
  network for anisotropic elastoplastic materials},
\newblock \bibinfo{journal}{Computer Methods in Applied Mechanics and
  Engineering} \bibinfo{volume}{363} (\bibinfo{year}{2020})
  \bibinfo{pages}{112875}.
  \DOIprefix\doi{https://doi.org/10.1016/j.cma.2020.112875}.
\bibitem[{Wu and Noels(2022)}]{WU2022114476}
\bibinfo{author}{L.~Wu}, \bibinfo{author}{L.~Noels},
\newblock \bibinfo{title}{Recurrent neural networks (rnns) with dimensionality
  reduction and break down in computational mechanics; application to
  multi-scale localization step},
\newblock \bibinfo{journal}{Computer Methods in Applied Mechanics and
  Engineering} \bibinfo{volume}{390} (\bibinfo{year}{2022})
  \bibinfo{pages}{114476}.
  \DOIprefix\doi{https://doi.org/10.1016/j.cma.2021.114476}.
\bibitem[{Logarzo et~al.(2021)Logarzo, Capuano, and Rimoli}]{LOGARZO2021}
\bibinfo{author}{H.~J. Logarzo}, \bibinfo{author}{G.~Capuano},
  \bibinfo{author}{J.~J. Rimoli},
\newblock \bibinfo{title}{Smart constitutive laws: Inelastic homogenization
  through machine learning},
\newblock \bibinfo{journal}{Computer Methods in Applied Mechanics and
  Engineering} \bibinfo{volume}{373} (\bibinfo{year}{2021})
  \bibinfo{pages}{113482}.
  \DOIprefix\doi{https://doi.org/10.1016/j.cma.2020.113482}.
\bibitem[{Gorji et~al.(2020)Gorji, Mozaffar, Heidenreich, Cao, and
  Mohr}]{GORJI2020}
\bibinfo{author}{M.~B. Gorji}, \bibinfo{author}{M.~Mozaffar},
  \bibinfo{author}{J.~N. Heidenreich}, \bibinfo{author}{J.~Cao},
  \bibinfo{author}{D.~Mohr},
\newblock \bibinfo{title}{On the potential of recurrent neural networks for
  modeling path dependent plasticity},
\newblock \bibinfo{journal}{Journal of the Mechanics and Physics of Solids}
  \bibinfo{volume}{143} (\bibinfo{year}{2020}) \bibinfo{pages}{103972}.
  \DOIprefix\doi{https://doi.org/10.1016/j.jmps.2020.103972}.
\bibitem[{Mozaffar et~al.(2019)Mozaffar, Bostanabad, Chen, Ehmann, Cao, and
  Bessa}]{Mozaffar26414}
\bibinfo{author}{M.~Mozaffar}, \bibinfo{author}{R.~Bostanabad},
  \bibinfo{author}{W.~Chen}, \bibinfo{author}{K.~Ehmann},
  \bibinfo{author}{J.~Cao}, \bibinfo{author}{M.~A. Bessa},
\newblock \bibinfo{title}{Deep learning predicts path-dependent plasticity},
\newblock \bibinfo{journal}{Proceedings of the National Academy of Sciences}
  \bibinfo{volume}{116} (\bibinfo{year}{2019}) \bibinfo{pages}{26414--26420}.
  \DOIprefix\doi{10.1073/pnas.1911815116}.
  \href{http://arxiv.org/abs/https://www.pnas.org/content/116/52/26414.full.pdf}{{\tt
  arXiv:https://www.pnas.org/content/116/52/26414.full.pdf}}.
\bibitem[{Koeppe et~al.(2021)Koeppe, Bamer, Selzer, Nestler, and
  Markert}]{Koeppe2021}
\bibinfo{author}{A.~Koeppe}, \bibinfo{author}{F.~Bamer},
  \bibinfo{author}{M.~Selzer}, \bibinfo{author}{B.~Nestler},
  \bibinfo{author}{B.~Markert}, \bibinfo{title}{Explainable artificial
  intelligence for mechanics: physics-informing neural networks for
  constitutive models}, \bibinfo{year}{2021}.
  \href{http://arxiv.org/abs/2104.10683}{{\tt arXiv:2104.10683}}.
\bibitem[{Liu et~al.(2023)Liu, Ocegueda, Trautner, Stuart, and
  Bhattacharya}]{LIU2023105329}
\bibinfo{author}{B.~Liu}, \bibinfo{author}{E.~Ocegueda},
  \bibinfo{author}{M.~Trautner}, \bibinfo{author}{A.~M. Stuart},
  \bibinfo{author}{K.~Bhattacharya},
\newblock \bibinfo{title}{Learning macroscopic internal variables and history
  dependence from microscopic models},
\newblock \bibinfo{journal}{Journal of the Mechanics and Physics of Solids}
  \bibinfo{volume}{178} (\bibinfo{year}{2023}) \bibinfo{pages}{105329}.
  \DOIprefix\doi{https://doi.org/10.1016/j.jmps.2023.105329}.
\bibitem[{Ghane et~al.(2023)Ghane, Fagerström, and
  Mirkhalaf}]{ghane2023recurrent}
\bibinfo{author}{E.~Ghane}, \bibinfo{author}{M.~Fagerström},
  \bibinfo{author}{M.~Mirkhalaf}, \bibinfo{title}{Recurrent neural networks and
  transfer learning for elasto-plasticity in woven composites},
  \bibinfo{year}{2023}. \href{http://arxiv.org/abs/2311.13434}{{\tt
  arXiv:2311.13434}}.
\bibitem[{Cheung and Mirkhalaf(2024)}]{CHEUNG2024}
\bibinfo{author}{H.~L. Cheung}, \bibinfo{author}{M.~Mirkhalaf},
\newblock \bibinfo{title}{A multi-fidelity data-driven model for highly
  accurate and computationally efficient modeling of short fiber composites},
\newblock \bibinfo{journal}{Composites Science and Technology}
  \bibinfo{volume}{246} (\bibinfo{year}{2024}) \bibinfo{pages}{110359}.
  \URLprefix
  \url{https://www.sciencedirect.com/science/article/pii/S0266353823004530}.
  \DOIprefix\doi{https://doi.org/10.1016/j.compscitech.2023.110359}.
\bibitem[{Haghighat et~al.(2021)Haghighat, Raissi, Moure, Gomez, and
  Juanes}]{HAGHIGHAT2021}
\bibinfo{author}{E.~Haghighat}, \bibinfo{author}{M.~Raissi},
  \bibinfo{author}{A.~Moure}, \bibinfo{author}{H.~Gomez},
  \bibinfo{author}{R.~Juanes},
\newblock \bibinfo{title}{A physics-informed deep learning framework for
  inversion and surrogate modeling in solid mechanics},
\newblock \bibinfo{journal}{Computer Methods in Applied Mechanics and
  Engineering} \bibinfo{volume}{379} (\bibinfo{year}{2021})
  \bibinfo{pages}{113741}.
  \DOIprefix\doi{https://doi.org/10.1016/j.cma.2021.113741}.
\bibitem[{Arora et~al.(2022)Arora, Kakkar, Dey, and Chakraborty}]{Arora2022}
\bibinfo{author}{R.~Arora}, \bibinfo{author}{P.~Kakkar},
  \bibinfo{author}{B.~Dey}, \bibinfo{author}{A.~Chakraborty},
  \bibinfo{title}{Physics-informed neural networks for modeling rate- and
  temperature-dependent plasticity}, \bibinfo{year}{2022}.
  \href{http://arxiv.org/abs/2201.08363}{{\tt arXiv:2201.08363}}.
\bibitem[{Masi and Stefanou(2022)}]{Masi2021}
\bibinfo{author}{F.~Masi}, \bibinfo{author}{I.~Stefanou},
\newblock \bibinfo{title}{Multiscale modeling of inelastic materials with
  thermodynamics-based artificial neural networks (tann)},
\newblock \bibinfo{journal}{Computer Methods in Applied Mechanics and
  Engineering} \bibinfo{volume}{398} (\bibinfo{year}{2022})
  \bibinfo{pages}{115190}.
  \DOIprefix\doi{https://doi.org/10.1016/j.cma.2022.115190}.
\bibitem[{Eghbalian et~al.(2023)Eghbalian, Pouragha, and Wan}]{EGHBALIAN2023}
\bibinfo{author}{M.~Eghbalian}, \bibinfo{author}{M.~Pouragha},
  \bibinfo{author}{R.~Wan},
\newblock \bibinfo{title}{A physics-informed deep neural network for surrogate
  modeling in classical elasto-plasticity},
\newblock \bibinfo{journal}{Computers and Geotechnics} \bibinfo{volume}{159}
  (\bibinfo{year}{2023}) \bibinfo{pages}{105472}.
  \DOIprefix\doi{https://doi.org/10.1016/j.compgeo.2023.105472}.
\bibitem[{Garanger et~al.(2023)Garanger, Kraus, and
  Rimoli}]{garanger2023symmetryenforcing}
\bibinfo{author}{K.~Garanger}, \bibinfo{author}{J.~Kraus},
  \bibinfo{author}{J.~J. Rimoli}, \bibinfo{title}{Symmetry-enforcing neural
  networks with applications to constitutive modeling}, \bibinfo{year}{2023}.
  \href{http://arxiv.org/abs/2312.13511}{{\tt arXiv:2312.13511}}.
\bibitem[{Wen et~al.(2021)Wen, Zou, and Wei}]{WEN2021}
\bibinfo{author}{J.~Wen}, \bibinfo{author}{Q.~Zou}, \bibinfo{author}{Y.~Wei},
\newblock \bibinfo{title}{Physics-driven machine learning model on temperature
  and time-dependent deformation in lithium metal and its finite element
  implementation},
\newblock \bibinfo{journal}{Journal of the Mechanics and Physics of Solids}
  \bibinfo{volume}{153} (\bibinfo{year}{2021}) \bibinfo{pages}{104481}.
  \DOIprefix\doi{https://doi.org/10.1016/j.jmps.2021.104481}.
\bibitem[{Bhattacharya et~al.(2023)Bhattacharya, Liu, Stuart, and
  Trautner}]{Bhattacharya2023}
\bibinfo{author}{K.~Bhattacharya}, \bibinfo{author}{B.~Liu},
  \bibinfo{author}{A.~Stuart}, \bibinfo{author}{M.~Trautner},
\newblock \bibinfo{title}{Learning markovian homogenized models in
  viscoelasticity},
\newblock \bibinfo{journal}{Multiscale Modeling \& Simulation}
  \bibinfo{volume}{21} (\bibinfo{year}{2023}) \bibinfo{pages}{641--679}.
  \DOIprefix\doi{10.1137/22M1499200}.
\bibitem[{Ge and Tagarielli(2021)}]{Ge2021}
\bibinfo{author}{W.~Ge}, \bibinfo{author}{V.~L. Tagarielli},
\newblock \bibinfo{title}{A computational framework to establish data-driven
  constitutive models for time- or path-dependent heterogeneous solids},
\newblock \bibinfo{journal}{Scientific Reports} \bibinfo{volume}{11}
  (\bibinfo{year}{2021}). \DOIprefix\doi{10.1038/s41598-021-94957-0}.
\bibitem[{Ghavamian and Simone(2019)}]{Ghavamian2019}
\bibinfo{author}{F.~Ghavamian}, \bibinfo{author}{A.~Simone},
\newblock \bibinfo{title}{{Accelerating multiscale finite element simulations
  of history-dependent materials using a recurrent neural network}},
\newblock \bibinfo{journal}{Computer Methods in Applied Mechanics and
  Engineering} \bibinfo{volume}{357} (\bibinfo{year}{2019})
  \bibinfo{pages}{112594}. \DOIprefix\doi{10.1016/j.cma.2019.112594}.
\bibitem[{Chen(2021)}]{Chen2021}
\bibinfo{author}{G.~Chen},
\newblock \bibinfo{title}{Recurrent neural networks ({RNNs}) learn the
  constitutive law of viscoelasticity},
\newblock \bibinfo{journal}{Computational Mechanics} \bibinfo{volume}{67}
  (\bibinfo{year}{2021}) \bibinfo{pages}{1009--1019}.
  \DOIprefix\doi{10.1007/s00466-021-01981-y}.
\bibitem[{Eghtesad et~al.(2023)Eghtesad, Fuhg, and Bouklas}]{Eghtesad2023}
\bibinfo{author}{A.~Eghtesad}, \bibinfo{author}{J.~N. Fuhg},
  \bibinfo{author}{N.~Bouklas}, \bibinfo{title}{Nn-evp: A physics informed
  neural network-based elasto-viscoplastic framework for predictions of grain
  size-aware flow response under large deformations}, \bibinfo{year}{2023}.
  \href{http://arxiv.org/abs/2307.04301}{{\tt arXiv:2307.04301}}.
\bibitem[{Maia et~al.(2023)Maia, Rocha, Kerfriden, and {van der
  Meer}}]{Maiaetal2023}
\bibinfo{author}{M.~Maia}, \bibinfo{author}{I.~Rocha},
  \bibinfo{author}{P.~Kerfriden}, \bibinfo{author}{F.~{van der Meer}},
\newblock \bibinfo{title}{Physically recurrent neural networks for
  path-dependent heterogeneous materials: Embedding constitutive models in a
  data-driven surrogate},
\newblock \bibinfo{journal}{Computer Methods in Applied Mechanics and
  Engineering} \bibinfo{volume}{407} (\bibinfo{year}{2023})
  \bibinfo{pages}{115934}.
  \DOIprefix\doi{https://doi.org/10.1016/j.cma.2023.115934}.
\bibitem[{Rocha et~al.(2023)Rocha, Kerfriden, and {van der
  Meer}}]{Rochaetal2023}
\bibinfo{author}{I.~Rocha}, \bibinfo{author}{P.~Kerfriden},
  \bibinfo{author}{F.~{van der Meer}},
\newblock \bibinfo{title}{Machine learning of evolving physics-based material
  models for multiscale solid mechanics},
\newblock \bibinfo{journal}{Mechanics of Materials} \bibinfo{volume}{184}
  (\bibinfo{year}{2023}) \bibinfo{pages}{104707}.
  \DOIprefix\doi{https://doi.org/10.1016/j.mechmat.2023.104707}.
\bibitem[{Kovačević and {van der Meer}(2022)}]{Kovacevic2022}
\bibinfo{author}{D.~Kovačević}, \bibinfo{author}{F.~P. {van der Meer}},
\newblock \bibinfo{title}{Strain-rate based arclength model for nonlinear
  microscale analysis of unidirectional composites under off-axis loading},
\newblock \bibinfo{journal}{International Journal of Solids and Structures}
  \bibinfo{volume}{250} (\bibinfo{year}{2022}) \bibinfo{pages}{111697}.
  \DOIprefix\doi{https://doi.org/10.1016/j.ijsolstr.2022.111697}.
\bibitem[{Bonet and Burton(1998)}]{Bonet1998}
\bibinfo{author}{J.~Bonet}, \bibinfo{author}{A.~Burton},
\newblock \bibinfo{title}{A simple orthotropic, transversely isotropic
  hyperelastic constitutive equation for large strain computations},
\newblock \bibinfo{journal}{Computer Methods in Applied Mechanics and
  Engineering} \bibinfo{volume}{162} (\bibinfo{year}{1998})
  \bibinfo{pages}{151--164}. \URLprefix
  \url{https://www.sciencedirect.com/science/article/pii/S0045782597003393}.
  \DOIprefix\doi{https://doi.org/10.1016/S0045-7825(97)00339-3}.
\bibitem[{Kingma and Ba(2014)}]{Adametal2014}
\bibinfo{author}{D.~P. Kingma}, \bibinfo{author}{J.~Ba}, \bibinfo{title}{Adam:
  A method for stochastic optimization}, \bibinfo{year}{2014}.
  \DOIprefix\doi{10.48550/ARXIV.1412.6980}.
\bibitem[{Chen and Wheeler(1993)}]{Chen1993}
\bibinfo{author}{Y.-C. Chen}, \bibinfo{author}{L.~Wheeler},
\newblock \bibinfo{title}{Derivatives of the stretch and rotation tensors},
\newblock \bibinfo{journal}{Journal of Elasticity} \bibinfo{volume}{32}
  (\bibinfo{year}{1993}) \bibinfo{pages}{175--182}.
  \DOIprefix\doi{10.1007/bf00131659}.
\bibitem[{Rocha et~al.(2020)Rocha, Kerfriden, and van~der Meer}]{Rocha2020}
\bibinfo{author}{I.~B. Rocha}, \bibinfo{author}{P.~Kerfriden},
  \bibinfo{author}{F.~P. van~der Meer},
\newblock \bibinfo{title}{Micromechanics-based surrogate models for the
  response of composites: A critical comparison between a classical mesoscale
  constitutive model, hyper-reduction and neural networks},
\newblock \bibinfo{journal}{European Journal of Mechanics, A/Solids}
  \bibinfo{volume}{82} (\bibinfo{year}{2020}) \bibinfo{pages}{103995}.
  \DOIprefix\doi{10.1016/j.euromechsol.2020.103995}.

\end{thebibliography}

\end{document}